\pdfoutput=1
\documentclass[reqno]{amsart}
\usepackage[margin=1in]{geometry}
\usepackage{amsmath, amsthm, amssymb, amsfonts}

\usepackage[dvipsnames]{xcolor}
\usepackage{soul}
\usepackage{graphicx}
\usepackage{float}
\usepackage{amsmath,amssymb}
\newtheorem{rmk}{Remark}

\newcommand{\tmk}{\tanh(\mu k)}

\newcommand{\tmdk}{\tanh(\delta \mu k)}

\newcommand{\FTk}[1]{\mathcal{F}_k\left\lbrace #1 \right\rbrace}
\newcommand{\specInt}[1]{\frac{1}{2\pi}\int_0^{2\pi}e^{-ikx}  #1 \:dx}

\newcommand{\beq}{\begin{equation}}
\newcommand{\eeq}{\end{equation}}

\newcommand{\ba}{\begin{array}}
\newcommand{\ea}{\end{array}}

\newcommand{\bea}{\begin{eqnarray}}
\newcommand{\eea}{\end{eqnarray}}

\newcommand{\bc}{\begin{center}}
\newcommand{\ec}{\end{center}}

\newcommand{\bt}{\begin{tabular}}
\newcommand{\et}{\end{tabular}}

\newcommand{\bi}{\begin{itemize}}
\newcommand{\ei}{\end{itemize}}

\newcommand{\bd}{\begin{description}}
\newcommand{\ed}{\end{description}}

\newcommand{\bp}{\begin{pmatrix}}
\newcommand{\ep}{\end{pmatrix}}

\newcommand{\tkz}{\mathcal{T}_{k_0}}

\newcommand{\p}{\partial}

\title{Nonlinear Traveling Internal Waves in Depth-Varying Currents}

\author{K. L. Oliveras$^1$}
\address{$^1$Mathematics Department, Seattle University, Seattle, WA 98122}
\author{C. W. Curtis$^2$}
\address{$^2$Department of Mathematics and Statistics, San Diego State University, San Diego, CA 92182}

\begin{document}

\begin{abstract}
 In this work, we study the nonlinear traveling waves in density stratified fluids with depth varying shear currents.  Beginning the formulation of the water-wave problem due to \cite{afm}, we extend the work of \cite{ashton} and \cite{haut} to examine the interface between two fluids of differing densities and varying linear shear.  We derive as systems of equations depending only on variables at the interface, and numerically solve for periodic traveling wave solutions using numerical continuation.  Here we consider only branches which bifurcate from solutions where there is no slip in the tangential velocity at the interface for the trivial flow. The spectral stability of these solutions is then determined using a numerical Fourier-Floquet technique.  We find that the strength of the linear shear in each fluid impacts the stability of the corresponding traveling wave solutions.  Specifically, opposing shears may amplify or suppress instabilities. 
\end{abstract}
\maketitle

% \begin{keywords}
% Authors should not enter keywords on the manuscript, as these must be chosen by the author during the online submission process and will then be added during the typesetting process (see http://journals.cambridge.org/data/\linebreak[3]relatedlink/jfm-\linebreak[3]keywords.pdf for the full list)
% \end{keywords}

%\tableofcontents

% = = = = = = = = = = = = = = = = = = = = = = = = = = = = = = = = = = = = = = = = = = = = = = = = = = = = = = = = = = 
\section{Introduction}
% = = = = = = = = = = = = = = = = = = = = = = = = = = = = = = = = = = = = = = = = = = = = = = = = = = = = = = = = = = 
As noted in \cite{appel} and \cite{helfrich}, %building off the now seminal results from the remote sensing of coastal regions, 
internal waves are known to be a common feature in coastal oceanic flows.  Resulting from instabilities along pynoclines and thermoclines, internal waves with amplitudes from 5-100 m \cite{appel,helfrich,osborne} have been observed.  While several generation mechanisms have been proposed and studied, it appears that in most cases internal waves are generated in coastal regions due to the presence of tide induced shearing forces in stratified fluids moving over varying bathymetry \cite{farmer2,helfrich}.  What is interesting though is that while depth varying shear currents should be anticipated in such environments, most modeling attempts have traditionally ignored the impact of vorticity on internal wave dynamics.  This work will address this shortcoming in part by studying the impact of depth varying shear currents on the existence and flow properties of internal traveling wave solutions (TWSs) in density stratified environments.  

While the impact of vorticity on internal waves is not well understood, a great deal of work on the influence of shear in free surface flows has appeared.  The simplest case of free surfaces moving over depth varying currents is when the vorticity is assumed to be a constant throughout the fluid.  The literature on this subject is too vast to address here, though we point out the seminal studies \cite{dasilva,simmen,vandenbroeck} and the extensive bibliography in \cite{constantin}.  We also note that a key part of the approach we take in this paper follows the work in \cite{ashton}, which in part makes use of the Unified Transform Method (UTM) pioneered in \cite{fokas2008}.  

Perhaps surprisingly, there are still outstanding and significant issues even for constant vorticity flows.  In particular, the impact of vorticity on the stability of traveling wave solutions is not fully understood.  As shown in \cite{sprenger}, the impact of vorticity modifies the amplitudes of Benjamin--Feir (BF) instabilities depending on the sign of the vorticity.  Similar results have analytically been found in approximate models  \cite{choi2009nonlinear,hur2015modulational,thomas2012nonlinear}.  In addition, constant vorticity introduces ``high-frequency" instabilities similar to those observed  in irrotational waves \cite{McLean2,franciuskharif,oliveras}.  These instabilities may be more dominant than the BF instability depending on the amplitude of the TWS and the strength of the linear shear \cite{sprenger}.  Furthermore, for a TWS with a region of lower pressure trapped under the wave \cite{dasilva,vasan2014pressure}, new instabilities, perhaps related to a Rayleigh-Taylor instability, are visible in the spectrum.   As shown in \cite{vasan2014pressure}, the existence of these low pressure regions is directly related to the existence of stagnation points within the fluid.  While usually avoided in much of the existing literature due to the analytic difficulties they introduce, it is becoming clearer that stagnation points can have significant impacts on wave shape and stability.  Thus, in part due to stagnation point induced low pressure regions, and in part due to high-frequency instabilities, no TWSs have been found which are stable to all perturbations despite the ability of a constant shear current to suppress the onset of BF instabilities.   

%And beyond TWSs, what other dynamics might we observe if we allow for depth varying vorticity?

\begin{figure}
\centering
\includegraphics[width=\textwidth]{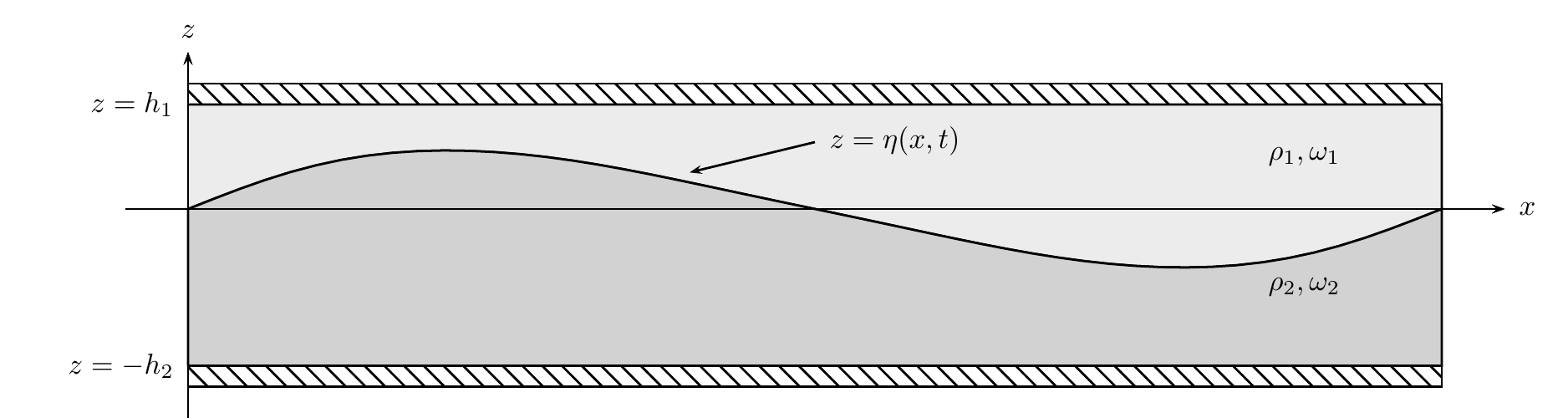}
\caption{Fluid Domain for the rigid lid problem}
\label{fig:fluid_flows_shear}
\end{figure}

Given these results for constant vorticity, it is interesting then to ask how depth variation in the vorticity affects the existence of these instabilities, and whether more complicated shear profiles will support the existence of stable internal or even free surface traveling waves.  The case of non-constant vorticity flows has received far less attention.  For the case of non-steady waves, this was first examined in \cite{dalrymple} for a stratified fluid with two differing densities $\rho_1$ and  $\rho_2$ and two differing constant vortices $\omega_1$ and $\omega_2$.  We refer to such fluids as {\it bistratified}; see Figure \ref{fig:fluid_flows_shear} for reference.  More recently, such systems have been looked at experimentally and numerically in \cite{swan,ko}.   Assuming a rigid lid and zero vorticity, i.e. $\omega_{2}=0$, in the lower fluid region, nonlinear, large amplitude traveling interfacial waves for bistratified fluids were studied in \cite{pullin2,pullin4}.  A complete characterization of the linear stability of the small amplitude limit of these waves was presented in \cite{pullin3}.  Generalizing this work, the evolution of a free surface and a free internal layer for a bistratified fluid in a shallow-water approximation was studied in \cite{curtis3}.  Significant nonlinear phenomena like internal dispersive shock waves as a result of bistratification was reported for the first time.
		
Focusing then on the case of single-valued internal layers, by assuming a rigid lid at the fluid-air interface and an impermeable sea bed,  using the UTM techniques in \cite{haut,ashton} and incorporating techniques developed for traveling wave solutions as found in \cite{oliverasvasan}, we derive a closed system describing the evolution of the free internal surface in a bistratified fluid.  Using this formulation, we can analytically determine at which wave speeds bifurcations from flat states will occur.  By developing a Stokes expansion at this bifurcation point, we can analytically determine the behavior of the bifurcation curve in terms of wave height and speed in a small amplitude limit, thereby allowing for a better understanding of when stagnation points in the fluid velocity appear and when turning points in bifurcation curves appear.    

We then use numerical continuation to generate bifurcation curves which provide the speed-amplitude relationship for the traveling nonlinear interfacial waves.  Of particular interest is the role different choices of vorticities play in determining the number of stagnation points within the fluid regions, and how these stagnation points influence the shape and properties of the interfacial TWSs.  As we show, the position and number of stagnation points determines whether waves are elevated, depressed, symmetric in nature, or especially peaked.  To complement these results, we also present plots of the streamlines to help illustrate the different impacts on dynamics that the stagnation points have throughout the bulk of the fluid as well as on the interface.  

Finally, we study the stability of various TWSs.  One of the most striking features associated with bistratification is the complicated relationship between the appearance of modulational instabilities (MIs) and the choices of the shear strength and depth ratio of the layers.  As we show, by making the ratio of the depths of the two layers large enough, MIs appear to be completely suppressed.  However, we also show results in which MIs are introduced and suppressed by shear, thereby illustrating the complicated balances mediated by nonlinearity between these parameter choices.  These results generalize and expand on those for constant vorticity in single layer fluids \cite{thomas2012nonlinear,sprenger}, in which vorticity is shown to strongly influence whether MIs exist.

However, at least for the solutions examined in this paper, all traveling solutions are unstable with respect to  higher-frequency perturbations.  Given the more complicated nature of the physical problem studied in this paper, it is a non-trivial question to determine if all traveling wave solutions are unstable, especially given the stabilizing mechanisms we illustrate via our numerical result.  Likewise, the complex nature of instabilities for multi-valued interfaces remain unknown and will be explored in future work.

The structure of the paper is as follows.  The derivation of our model is given in Section 2.  The computation of the TWSs and analytic derivations of the nonlinear impact on wave speed and the presence of stagnation points are given in Section 3.  Section 4 explains the means by which we numerically determine the stability of the TWSs computed in this paper, and Section 5 contains the results of these stability computations.  Concluding remarks appear in Section 6, and an Appendix collections technical details used throughout the paper.  

% = = = = = = = = = = = = = = = = = = = = = = = = = = = = = = = = = = = = = = = = = = = = = = = = = = = = = = = = = = 
\section{Boundary Conditions and AFM Formulation}
We assume the flow is incompressible in each region in a frame of reference moving to the right with speed $c$.  The resulting equations must satisfy Euler's equations in the bulk:
\begin{align}
	&\nabla\cdot\mathbf{u}_j =0,\label{eqn:conMass}\\
	&\partial_t\mathbf{u}_j+(\mathbf{u}_j\cdot\nabla)\mathbf{u}_j = -\frac{1}{\rho_{j}}\nabla{p}_j	- g {\bf{\hat{z}}},\label{eqn:conMom}
\end{align}
where ${\bf u}_{j}=(u_{j}-c,w_{j})^{T}$ and $j=1,2$ corresponds to the fluid velocities in the top and bottom layer of the fluid respectively, $p_j = p_j(x,z)$ is the pressure at any point within each fluid, $\rho_j$ is the density of each fluid, and $g$ is the constant acceleration due to gravity. 

At the interface between each fluid, we require continuity in pressure, as well as continuity in the normal velocities of each fluid at the interface.  This provides the boundary conditions 
\begin{equation}
	p_1 = p_2, ~\textrm{and}~\mathbf{u}_1\cdot \mathbf{N} = \mathbf{u}_2\cdot \mathbf{N} \quad \text{at } z = \eta(x,t),\label{eqn:interfaceBCU}
	\end{equation}
where $\bf{N}$ represents a unit normal vector to the interface defined by $z = \eta(x,t)$.  Furthermore, the interface is evolved by the fluid velocities normal to the interface.  That is, $$\eta_t = \mathbf{u}_j\cdot\mathbf{N},$$ for $j = 1$ or $j = 2$ as they are equivalent.  Finally, the rigid lid and solid boundary conditions require 
\begin{equation}
	v_1 =0 ~\text{at } z = h_1, \qquad v_2 = 0 ~\text{at }z = -h_2.\label{eqn:noFluxU}
	\end{equation}

\subsection{A Non-local Formulation}

Equations (\ref{eqn:conMass}-\ref{eqn:noFluxU}) represent the equations we wish to solve provided that there is a constant linear shear within each fluid domain.  That is, that $$\partial_xw_j - \partial_zu_j = \omega_j$$ within each fluid where we have chosen a particular sign convention for the vorticity consistent with conventions used elsewhere in the literature (see \cite{dasilva,ko} for example).  A disadvantage of the above formulation is that it requires one to solve for the velocities and pressure within the bulk of each fluid as well as the interface separating the two fluids.  In our work, we choose to cast the equations of motion using a nonlocal reformulation that allows us to work completely within the context of variable defined at the interface.  Here we following the work of \cite{ashton,haut} in order to recast the problem into variables defined at the interface.  However, to take advantage of this formulation, we first pose (\ref{eqn:conMass}-\ref{eqn:noFluxU}) in terms of new bulk variables.

Let $\psi_j$ for $j = 1, 2$ represent the stream functions in the top and bottom layer of the fluid respectively.  Furthermore, we introduce the functions $\phi_j$ for $j = 1, 2$ to represent the contributions to the velocity from potential flow.  Thus, we can write the fluid velocities  as 
\[
\bp u_{j} -c \\ w_{j}\ep = \bp \partial_x\phi_j -\omega_j z\\ \partial_z\phi_j\ep = \bp \partial_z\psi_j\\-\partial_x\psi_j\ep.
\]
Using \eqref{eqn:conMass}-\eqref{eqn:noFluxU}, the stream- and potential-functions satisfy
\begin{align}
	&\Delta \phi_j = 0, \quad &&(x,z)\in D_j\label{eqn:laplacePhi}\\
	&\Delta \psi_j = -\omega_j\quad &&(x,z)\in D_j\label{eqn:poissonPsi},\\
	&\partial_t\phi_j + \omega_j\psi_j + \frac{1}{2}\vert\nabla\phi_j - \omega_jz\hat{\bf{x}}\vert^2 + \frac{p_j}{\rho_j} + g z = 0, &&(x,z)\in D_j\label{eqn:bulkBernoulli}
	\end{align}
within the bulk of each fluid.
 %Equations \eqref{eqn:laplacePhi}-\eqref{eqn:bulkBernoulli} represent the equations of motion within the bulk of each fluid.  
\begin{rmk}
	In integrating \eqref{eqn:conMom} within the fluid domain to find \eqref{eqn:bulkBernoulli}, we arrive at an equation of the form 
	$$\partial_t\phi_j + \omega_j\psi_j + \frac{1}{2}\vert\nabla\phi - \omega_jz\hat{\bf{x}}\vert^2 + \frac{p_j}{\rho_j} + g z = B_j(t)$$ where we call the time varying function $B_{j}(t)$ the Bernoulli function.  However, by introducing the change of variables
	\[
	\phi_{j} = \tilde{\phi}_{j} + \int_{0}^{t}B_{j}(s) ds,
	\]
	we can eliminate the Bernoulli function, and thus we set it to zero throughout the remainder of this text.
	\end{rmk}

\subsubsection{Boundary Conditions}  Examining the conditions prescribed at the interface given by \eqref{eqn:interfaceBCU}, we find
\[
\partial_z\phi_j - \left(\partial_x\phi_j - \omega_j\eta\right)\partial_x\eta =  \partial_t \eta.
\]
Likewise, we can readily show that 
\[
\frac{d}{dx}\left(\psi_j(x,\eta(x,t),t)\right) = -\partial_t \eta,
\]
so that 
\[
\psi_j(x,\eta(x,t),t) = -\partial_x^{-1}\partial_t\eta 
\] 
where we have picked the arbitrary constant for the domain.  	Invoking the pressure condition that $p_1 = p_2$ at the free interface, we then find 
\begin{multline*}
\partial_t\phi_1 - \omega_1\partial_x^{-1}\partial_t\:\eta + \frac{1}{2}\left(\left(\partial_x\phi_{1} - \omega_1\eta \right)^2 + \left(\partial_z\phi_{1}\right)^2\right)  + g\eta  \\
			= 	\rho \left(\partial_t\phi_2 - \omega_2\partial_x^{-1}\partial_t\:\eta + \frac{1}{2}\left(\left(\partial_x\phi_{2} - \omega_2\eta \right)^2 + \left(\partial_z\phi_{2}\right)^2\right)  + g\eta \right)
	\end{multline*}
where $\rho = \rho_{2}/\rho_{1}$.  Finally, at the rigid lid and the solid bottom boundary, we have the zero flux conditions which require 
\[
\partial_z\phi_1 = 0  \text{ at }  z=h_1,~\text{and }~ \partial_z\phi_2 = 0 \text{ at } -h_2.
\]

\subsection{Nonlocal Formulation in Interface Quantities}
	%\hl{So that we don't have to do the ``avg value on the whole line'' in the stability section, we might just consider going with that directly here.  }

	Following the work of \cite{afm,ashton,haut}, we introduce the harmonic functions $\varphi_j$ such that
	\[ \varphi_{j} = e^{-ikx}\cosh\left(k(z+(-1)^{j}h_{j})\right).\] 
	Noting that $$(\partial_z\varphi_j) \left(\Delta\psi_j+\omega_j\right) + \partial_z\psi_j\left(\Delta\varphi_j\right) = 0,$$ where $\psi_j$ solves \eqref{eqn:poissonPsi}, we integrate the above quantity over the respective fluid domain.  By using Green's second identity and integration by parts, we can readily show that for $j = 1, 2$.
	
	\begin{equation}
		\int_{0}^{2\pi L} e^{-ikx}\left(\left(ik(\partial_t\eta)- \omega_j \right)\mathcal{C}_{j} - k\left(\partial_{x}q_{j} - \omega_j\eta \right)\mathcal{S}_{j}\right) dx = 0, \forall k \in \Lambda\label{eqn:nonlocalInt}
		\end{equation}
	where $\Lambda = \left\lbrace k = \frac{2n\pi}{L}\vert n \in \mathbb{Z}/\lbrace 0 \rbrace\right\rbrace$, $q_j$ is given by \[q_{j}(x,t) = \phi_{j}(x,\eta(x,t),t),\] and  the quantities $\mathcal{C}_j$, $\mathcal{S}_j$ are defined by
	\[\mathcal{C}_{j} = \cosh\left(k(\eta+(-1)^{j}h_{j})\right), \quad \mathcal{S}_{j} = \sinh\left(k(\eta+(-1)^{j}h_{j})\right). \] 
	
	Using the definition for $q_j$, we can rewrite both  $\partial_x\phi_j$ and $\partial_z\phi_j$ both evaluated at $z = \eta$ in terms of our new variables $q_1$ and $q_2$ as
	$$\partial_x\phi_j  - \omega_j \eta = \frac{\partial_xq_j  - \omega_j\eta - \partial_x\eta\partial_t\eta}{1 + \left(\partial_x\eta\right)^2}, \qquad \partial_z\phi_j = \frac{\partial_t\eta+\partial_x\eta\left(\partial_xq_j  - \omega_j\eta\right)}{1 + \left(\partial_x\eta\right)^2}.$$%
	Similarly, we can also write $\partial_t\phi_j$ as $$\partial_t\phi_j = \partial_tq_j - \frac{\partial_t\eta\left(\partial_t\eta + \partial_x\eta\left(\partial_xq_j - \omega_j\eta\right)\right)}{1 + \left(\partial_x\eta\right)^2}$$
	Using the boundary condition $$p_1(x,\eta,t) = p_2(x,\eta,t),$$ along with our new variables $q_j$ for $j = 1,2$, we have
	\begin{multline}
	\partial_tq_1 - \omega_1\partial_x^{-1}\partial_t\eta + \frac{1}{2}\left(\partial_xq_1 - \omega_1\eta\right)^2 - \frac{1}{2}\frac{\left(\partial_t\eta + \partial_x\eta\left(\partial_xq_1 - \omega_1\eta\right)\right)^2}{1 + \left(\partial_x\eta\right)^2} + g\eta  \\
	 = \rho \left(\partial_tq_2 - \omega_2\partial_x^{-1}\partial_t\eta + \frac{1}{2}\left(\partial_xq_2 - \omega_2\eta\right)^2 - \frac{1}{2}\frac{\left(\partial_t\eta + \partial_x\eta\left(\partial_xq_2 - \omega_2\eta\right)\right)^2}{1 + \left(\partial_x\eta\right)^2} + g\eta\right).\label{eqn:dimBernoulli}
	\end{multline}
	Thus, \eqref{eqn:nonlocalInt} for $j = 1, 2$ along with \eqref{eqn:dimBernoulli} represent a closed system of three equations for the three unknown quantities $(q_1, q_2, \eta)$.  

	\subsection{Non-Dimensionalization}
	Introducing the non-dimensional parameters given by 
	\begin{align*}
	\tilde{x} = \frac{x}{L}, \qquad \tilde{z} = \frac{z}{h_1}, \qquad \tilde{t} = \frac{c_0}{L} t, \qquad q_{j} =  \epsilon c_0 L\tilde{q}_{j}, \qquad \eta  = a\tilde{\eta}, \qquad \tilde k = L k,
	\end{align*}
	and
	\[
	\tilde{\omega}_{j} =  \omega_{j}\frac{h_1}{c_0}, \qquad \tilde{c} = \frac{c}{c_0}, \qquad  \mu = \frac{h_1}{L}, \qquad  \epsilon = \frac{a}{h_1}, \qquad  \delta = \frac{h_2}{h_1},
	\]
	where $c_0 = \sqrt{gh_1}$, we have the following system of three equations

	\begin{multline}
		\partial_t\left( q_1 - \omega_1\partial_x^{-1}\eta\right) + \frac{\epsilon}{2}\left(\partial_xq_1 - \omega_1\eta\right)^2 +\eta -  \frac{\epsilon\mu^2}{2}\frac{\left(\partial_t \eta + (\epsilon\partial_x\eta)\left(\partial_xq_1 - \omega_1\eta\right)\right)^2}{1 + \left(\epsilon\mu\partial_x\eta\right)^2}\\
		= \rho\left(\partial_t\left(q_2 - \omega_2\partial_x^{-1}\eta\right) + \frac{\epsilon}{2}\left(\partial_xq_2-\omega_2\eta \right)^2 +\eta -  \frac{\epsilon\mu^2}{2}\frac{\left(\partial_t \eta + (\epsilon\partial_x\eta)\left(\partial_xq_2 - \omega_2\eta\right)\right)^2}{1 + \left(\epsilon\mu\partial_x\eta\right)^2}\right),\label{eqn:bernoulliNonDim}
		\end{multline}
			%	\begin{multline}
				%\partial_t\left(q_1\right) + \frac{\epsilon}{2}\left(\partial_xq_1  - \omega_1\eta\right)^2 - \frac{\epsilon^2\mu^2}{2}\frac{\left(\partial_t\eta + \partial_x\eta (\partial_x q_1  - \epsilon\omega_1\eta) \right)^2}{1 + \left(\epsilon\mu\partial_x\eta\right)^2} + \eta  \\
			%				= \rho\left( \partial_t\left(q_2 - \omega_2\partial_x^{-1}\eta\right) + \frac{\epsilon}{2}\left(\partial_xq_2 - \omega_2\eta \right)^2 - \frac{\epsilon}{2}\frac{\left(\partial_t\eta + \epsilon\mu\partial_x\eta (\partial_xq_2  - \omega_2\eta)\right)^2}{1 + \left(\epsilon\mu\partial_x\eta\right)^2} + \eta \right),
			%	\end{multline}
	\begin{equation}
	\int_{0}^{2\pi} e^{-ikx}\left(\left(i k \epsilon\mu^2\partial_t\eta- \omega_1\right)\mathcal{C}_{1} -  \epsilon\mu k \left(\p_{x}q_{1} - \omega_1\eta\right)\mathcal{S}_1   \right) dx = 0, \quad \forall k \in \mathbb{Z}/\lbrace 0 \rbrace,\label{eqn:TopNonlocalNonDim}
		\end{equation}
	and 
	\begin{equation}
		\int_{0}^{2\pi} e^{-ikx}\left(\left(i k \epsilon\mu^2\partial_t\eta- \omega_2\right)\mathcal{C}_{2} -  \epsilon\mu k \left(\p_{x}q_{2} - \omega_2\eta\right)\mathcal{S}_2   \right) dx = 0, \quad \forall k \in \mathbb{Z}/\lbrace 0 \rbrace,\label{eqn:BottomNonlocalNonDim}
		\end{equation}

% = = = = = = = = = = = = = = = = = = = = = = = = = = = = = = = = = = = = = = = = = = = = = = = = = = = = = = = = = = 
\section{Traveling Wave Equations}
% = = = = = = = = = = = = = = = = = = = = = = = = = = = = = = = = = = = = = = = = = = = = = = = = = = = = = = = = = = 
Working with the nondimensional equations in a traveling coordinate frame, letting $\p_t\eta\to0$ as well as $\p_tq_j\to0$ for $j = 1,2$, the equations for stationary solutions in a traveling frame become 
\[
\left(\frac{1}{2}\frac{\left(\epsilon(\partial_xq_1  - \omega_1\eta)-c\right)^2}{1 + \left(\epsilon\mu\partial_x\eta\right)^2} + \epsilon\eta - \frac{1}{2}c^2\right)  = \rho\left(\frac{1}{2}\frac{\left(\epsilon(\partial_xq_2  - \omega_2\eta)-c\right)^2}{1 + \left(\epsilon\mu\partial_x\eta\right)^2} + \epsilon\eta- \frac{1}{2}c^2\right)
\]
\[
\int_{0}^{2\pi}e^{-ikx}\left(\mathcal{C}_1 \omega_1 + \mu k \mathcal{S}_1\left(\epsilon(\partial_xq_1 - \omega_1\eta) - c\right)\right)\:dx = 0.
\]
\[
\int_{0}^{2\pi}e^{-ikx}\left(\mathcal{C}_2 \omega_2 + \mu k \mathcal{S}_2\left(\epsilon(\p_x q_2 - \omega_2\eta) - c\right)\right)\:dx = 0.
\]
where $\mathcal{C}_j = \cosh(\mu k (\epsilon\eta+d_j))$ and $\mathcal{S}_j = \sinh(\mu k (\epsilon\eta + d_j))$ with $d_1 = -1$ and $d_2 = \delta$.  In the above fomulation, we are required to solve three equations for three unknown functions $\eta$, $q_1$ and $q_2$. However, as in \cite{oliveras,ashton}, we can reduce the dimensionality of the problem by formally solving the Bernoulli equation for either the quantity $\epsilon \partial_xq_1 - \epsilon\omega_1\eta - c$ or  $\epsilon \partial_xq_2 - \epsilon\omega_2\eta - c$.  Without loss of generality, we solve for the tangential velocity approaching the interface from above $\partial_xq_1$ in terms of $\partial_xq_2$ to find
	\begin{equation}
		\epsilon(\partial_xq_1  -\omega_1\eta) -c = \sigma \sqrt{\rho (\epsilon\partial_xq_2 - \epsilon\omega_2\eta - c)^2 + \left(c^2 - 2 \epsilon\eta\right)\left(1 + (\epsilon\mu\partial_x\eta)^2\right)\left(1-\rho\right)}\label{eqn:solve_q1x}
		\end{equation}
	where $\sigma = \pm 1$. 

\begin{rmk}
The meaning of the plus/minus sign is typically of critical importance in problems involving linear shear (see for example \cite{vasan2014pressure}).  However, when simply solving for the interface variables $\eta$ and $q_2$, an alternative choice in $\sigma$ is equivalent to changing the sign of vorticity in the upper layer given by $\omega_1$.  Furthermore, to ensure that $(\eta,q_1,q_2)=(0,0,0)$ remains a solution to the problem, that is, that we consider traveling wave solutions that bifurcate from flat water with no jump in the tangential velocity, we choose $\sigma = -\text{sgn}(c_0)$ as indicated by \eqref{eqn:solve_q1x} where $c_0$ is the speed corresponding to bifurcation from the trivial solution.  
\end{rmk}

The resulting equations to solve for $\eta(x)$ and $q_2(x)$ are given by 
\begin{multline}
\int_{0}^{2\pi}e^{-ikx}\left(\mathcal{C}_1 \omega_1 \right.\\+\left.\mu k\mathcal{S}_1\left(\sigma \sqrt{\rho (\epsilon\p_x q_2 - \epsilon\omega_2\eta - c)^2 + \left(c^2 - 2 \epsilon\eta\right)\left(1 + (\epsilon\mu\partial_x\eta)^2\right)\left(1-\rho\right)}\right)\right)\:dx = 0\label{eqn:nonlocalTWEA},
\end{multline}
\begin{equation}
\int_{0}^{2\pi}e^{-ikx}\left(\mathcal{C}_2 \omega_2 +\mu k\mathcal{S}_2\left(\epsilon(\partial_xq_2- \omega_2\eta) - c\right)\right)\:dx = 0,\label{eqn:nonlocalTWEB},
\end{equation}
where $\mathcal{C}_j$ and $\mathcal{S}_j$ are the same as given before  This system of two equations can be solved for the unknowns $\eta$ and $\partial_x q_2$ for an admissible value of the wave-speed $c$.  Once a solution $\eta$ and $\partial_xq_2$ are determined, the $\partial_x q_1$ is obtained directly by substitution in \eqref{eqn:solve_q1x}.

%-------------------------------------------------------------------------------------------------------------
%-------------------------------------------------------------------------------------------------------------
\subsection{Bifurcation from Trivial Flow}
%-------------------------------------------------------------------------------------------------------------
%-------------------------------------------------------------------------------------------------------------
Since we ultimately use a numerical continuation scheme in this work, we determine for which wave speeds $c$ a non-trivial solution will bifurcation from the zero solution.  This is achieved by expanding the free surface $\eta$, $\partial_xq_2$ and $c$ in powers of $\epsilon$ of the form 
\[
\eta(x) = \sum_{n = 0}^\infty \epsilon^n\eta_n(x),\qquad \partial_xq_2 = \sum_{n=0}^\infty \epsilon^n\partial_x v_n, \qquad c = \sum_{n=0}^\infty \epsilon^n c_n.
\] 
We assume that solutions will bifurcation from a wave with positive velocity so that $c_0>0$ and thus, choosing $\sigma = -1$ enforces bifurcating from a branch where there is no jump in the tangential velocity.  Expanding \eqref{eqn:nonlocalTWEA} and \eqref{eqn:nonlocalTWEB} in $\epsilon$, we find the following linear system for $\eta_0$ and $v_0$ given by 
\begin{equation}
\bp
\mu k  c_0^2 -\tmk\left(\tilde\omega c_0 - (\rho -1)\right) & c_0 \rho\tmk\\-\mu k c_0 & \tmdk
\ep \bp \FTk{\eta_0}\\\FTk{v_0}\ep = c_1\bp\FTk{ c_0\tmk}\\-\FTk{\tmdk}\ep,\label{eqn:order1LinSys}
\end{equation}
where $\FTk{f(x)}$ represents the $k$-th Fourier coefficient of $f(x)$ defined by $$\FTk{f(x)} = \specInt{f(x)},$$ and we have introduced the quantity $\tilde\omega = \omega_1 - \rho\omega_2.$
 	
 Since the right-hand side of \eqref{eqn:order1LinSys} evaluates to zero, we can then say that the only way for there to be a non-zero solution for $\eta_0$ and $v_0$ is for the linear operator on the left-hand side of \eqref{eqn:order1LinSys} singular for at least one $k$ value (henceforth referred to as $k_0$).  This leads to the equation $c_0$ given by 
\[
 c_0^2 - \tilde{\omega}\mathcal{T}_{k_0}c_0 - \mathcal{T}_{k_0}\left(\rho-1\right) = 0,
\]
where
\begin{equation} 
\mathcal{T}_k = \frac{\tmk\tmdk}{\mu k \left(\rho\tmk + \tmdk\right)}.\label{eqn:charEqn}
\end{equation} 
 	Using this notation, we find 
\begin{equation}
	c_0 = \frac{1}{2}\left(-\mathcal{T}_{k_0}\tilde{\omega} + \sqrt{\mathcal{T}_{k_0}^2\tilde{\omega}^2 + 4(\rho-1)\mathcal{T}_{k_0}}\right),\label{eqn:bifSpeed}
	\end{equation}
	 where the negative solution is extraneous due to our earlier assumptions of bifurcating from a positive speed $c_0$.  

The corresponding nontrivial solutions are given by 
\begin{equation}
\eta_0(x) = \cos(k_0 x), \qquad v_0(x) = \frac{\mu k_0 c_0}{\tanh(\mu\delta k_0)}\cos(k_0 x).\label{eqn:bifSol}
\end{equation}
Thus, for a given wave-number $k_0$, the leading-order approximations to the solution set $(\eta(x), \partial_xq_2,c)$ are given by the expressions in \eqref{eqn:bifSpeed} and \eqref{eqn:bifSol}.  With the aid of a CAS, we determine higher-order corrections both to the speed $c$ whereby
\[
c \approx c_0 + \epsilon^2 c_2 + \mathcal{O}(\epsilon^4).
\]
 Due to the complicated nature of the resulting expressions though, we obmit the details of computing $c_{2}$ here.  By finding $c_{2}$ though, we can generate analytic results which provide greater insight into the impact of vorticity on interfacial wave propagation.  These results will also help us identify potential regions of difficulty in the numerical results presented later in this paper.

\begin{figure}
\centering
\begin{tabular}{cc}
\includegraphics[width=.4\textwidth]{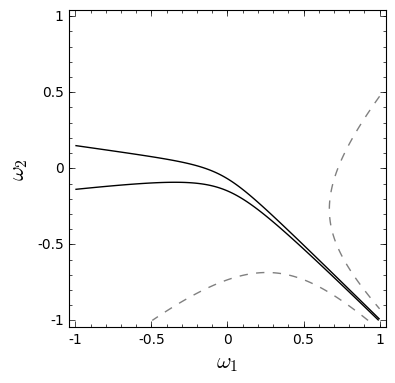} & %
\includegraphics[width=.4\textwidth]{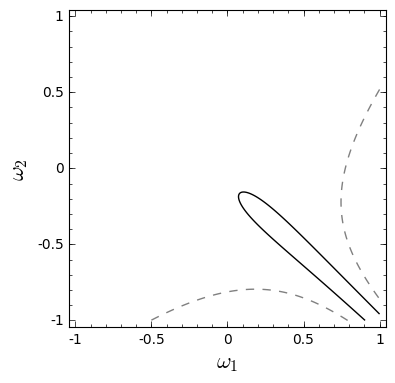}\\
\footnotesize{(a) $c_2$ for $k_0 = 1$ } & (b) \footnotesize{$c_2$ for $k_0=10$} 
\end{tabular}
\caption{Plots of $c_{2}$ for $\rho=1.028$, $\mu = \sqrt{.1}$, $\delta = 4$ and $k_0=1$ (a) and $k_{0}=10$ (b). In (a) and (b) the solid (--) line denotes the $c_{2}=0$ contour, with the interior of these curves marking the regions for which $c_{2}<0$.  In (a) the dashed curve (- -) denotes the $c_{2}=10^{6}$ contour, while in (b) it denotes the $c_{2}=10^{4}$ contour.}
\label{fig:lowrho}
\end{figure}

We present our results for computing $c_{2}$ in Figure \ref{fig:lowrho}.  As seen in the figure, the response of $c_{2}$ to the choice of vorticities can be quite sensitive.  For example, choosing a weak stratification value of $\rho=1.028$, reflective of oceanic density stratification, $\mu = \sqrt{.1}$, $\delta = 4$, and choosing the $c_{0,+}$ branch, we plot $c_{2}$ in Figure \ref{fig:lowrho} for $k_{0}=1$ and $k_{0}=10$.  While both plots exhibit a wide range of scales with respect to the magnitude of $c_{2}$, this range is far wider for $k_{0}=1$ than $k_{0}=10$.  Likewise, the region over which $c_{2}$ is negative is far larger in the case for $k_{0}=1$.  Note, those regions for which $c_{2}<0$ are regions in which the wave speed decreases with increasing wave height, which is in contrast to the usual result whereby nonlinear wave speed increases with increasing wave height.     

%By greatly increasing the density stratification to $\rho=820$, representing the density ratios of air to water, and inverting the depth ratio to $.25$, we get markedly different responses for $c_{2}$ as seen in Figure \ref{fig:highrho}.  In the case that $k_{0}=1$, we see in Figure \ref{fig:highrho} (a) that there is now no region in which $c_{2}<0$.  In contrast, for $k_{0}=10$, the entire lower portion of Figure \ref{fig:highrho} (b) is negative, and larger negative values are observed.  In both cases, the far stronger stratification and lower depth ratio reduces the sensitivity of $c_{2}$ to variations in $\omega_{j}$ relative to the sensitivity seen in the weak stratification case.  

\begin{figure}
\centering
\includegraphics[width=.5\textwidth]{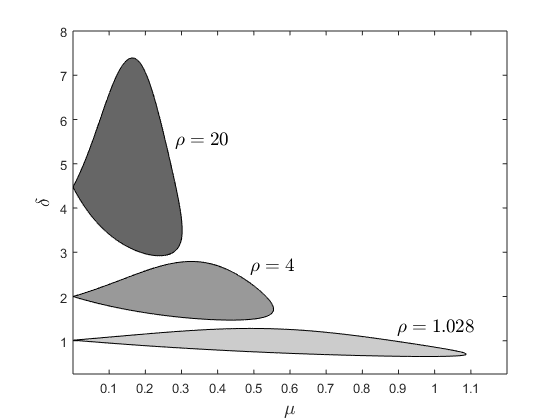}
\caption{For the irrotational case ($\omega_{1}=\omega_{2}=0$), regions where $c_2$ is negative as a function of $\mu$ and $\delta$ for discrete $\rho$ values ranging from $\rho = 1.028$ to $\rho = 20$.  The boundaries indicate parameter configurations for which $c_2 = 0$, and the interior corresponds to $c_2<0$.}\label{fig:turningPtMuvsDel}
\end{figure}

\subsection{Stagnation Points}
Given their connections to high pressure regions, the presence of stagnation points interior to each of the fluid domains impacts both the shape and stability of traveling wave solutions.  We proceed to find conditions for the presence of stagnation points interior to each of the fluid domains.  In order for there to be a stagnation point inside the fluid domain, there must exist a point inside the domain such that $u-c = 0$.  For simplicity, let $z_j^{(s)}$ represent the location of a stagnation point inside the upper ($j=1$) or lower ($j=2$) fluid domain so that $0\leq z_1^{(s)}\leq 1$ and $-\delta\leq z_2^{(s)}\leq 0$. At leading order, it is straightforward to show that the relative horizontal profile $u-c$ is given by 
\[
u - c = -\omega_j z - c_0 + \mathcal{O}(\epsilon)
\]
inside of each fluid domain implying that stagnation will occur if there is a $z$ inside each fluid domain such that $-\omega_j z =c_0$.  

\begin{figure}
\centering
\includegraphics[width=.5\textwidth]{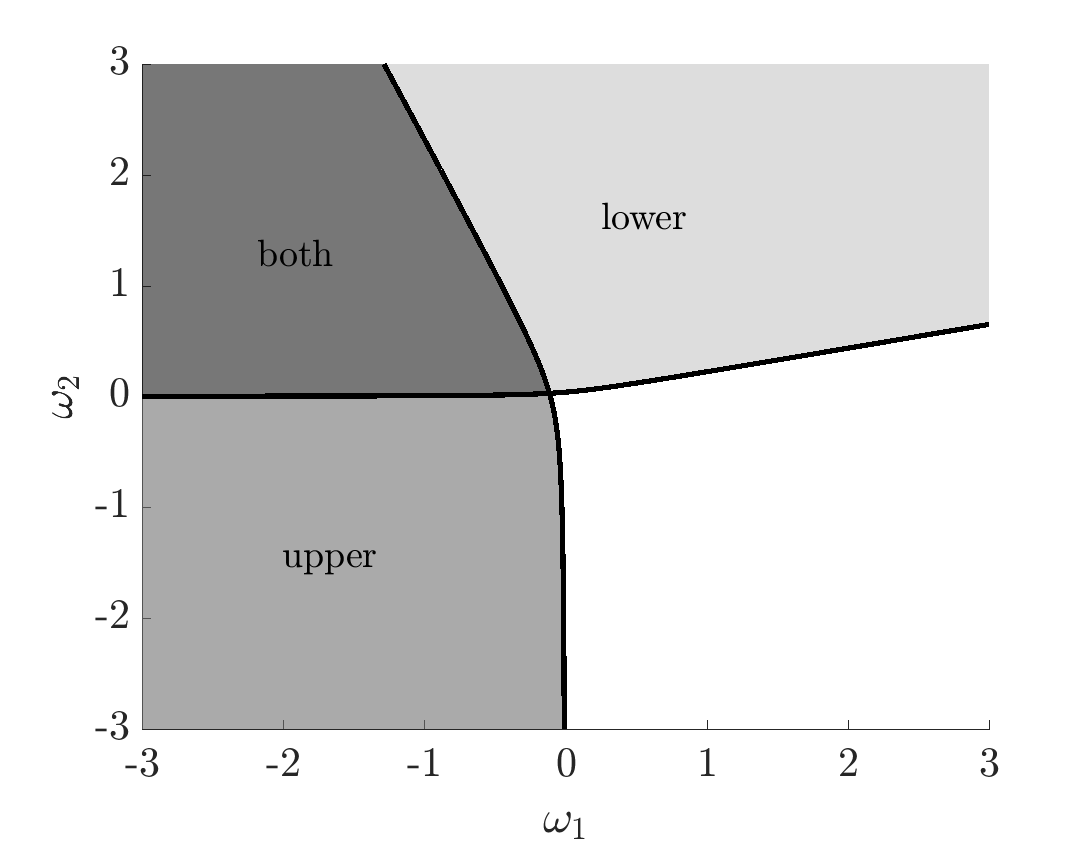}
\caption{Vorticity strengths that yield a stagnation point in the upper fluid, lower fluid, or in both for $\mu = \sqrt{.1}$, $\rho = 1.028$ and $\delta = 4$.}
\label{fig:stagnationPlotlowrho}
\end{figure}

For example, in the upper fluid domain, the condition for the existence of a stagnation point is given by  
\[
 -\omega_1z_1^{(s)} =c_0^{\pm} \qquad\to\qquad z_1^{(s)} = \frac{c_0^{\pm}}{-\omega_1}.
\]
From this perspective, it is clear that there will only be a stagnation point inside the upper fluid domain provided that $sgn(\omega_1) = -sgn(c_0^{\pm})$.  Likewise, there will only be a stagnation point inside the lower fluid domain provided that $sgn(\omega_2) = sgn(c_0^{\pm})$.  By enforcing the condition that the stagnation point is inside the appropriate fluid domain, we find the following conditions for the existence of a stagnation point  when $c_0 = c_0^+$: 
\begin{itemize}
\item[] \textit{Upper fluid domain:}	$\qquad \qquad\displaystyle \omega_1 > \frac{\rho\:\omega_2\tkz - \sqrt{\rho^2\omega_2^2\tkz^2 + 4\tkz(\tkz+1)(\rho-1)}}{2(\tkz+1)}.$\\~\\
\item[] \textit{Lower fluid domain:} $\qquad \qquad \displaystyle \omega_2 < \frac{\omega_1\delta \tkz + \sqrt{\omega_1^2\delta^2\tkz^2 + 4\delta\tkz(\rho\:\tkz+\delta)(\rho-1)}}{2\delta(\rho\tkz+\delta)}.$\\
\end{itemize}
Figure \ref{fig:stagnationPlotlowrho} show the regions in the $\omega_1,\omega_2$ plane where stagnations points are found in the upper, lower, or both fluid domains. Depending on the configuration of $\omega_1$ and $\omega_2$ the presence of the stagnation points for the trivial solution from which we continue greatly impact the shape of the solutions along the corresponding bifurcation branch.  These various configurations will be contrasted in the next section with traveling wave solutions.

%	\begin{figure}[htp]
%		\centering
%		\includegraphics[width=.5\textwidth]{./stagnation_delta_0_25_mu_0_31623_rho_820}
%		\caption{Vorticity strengths that yield a stagnation point for $\mu = 1$, $\rho = 820$ and $\delta = 0.25$.}
%		\label{fig:stagnationPlothighrho}
%		\end{figure}
%\begin{rmk} It is worthwhile to note that for $\omega_1 = -\omega_2$, as $\omega_1\to\infty$, there are stagnations points inside both fluid domains that limit towards the interface.  This can be constrasted with the negative corrections to $c_2$ shown in Figures \ref{fig:lowrho} - \ref{fig:highrho}.
%\end{rmk}
%\begin{figure}
%\centering
%\includegraphics[width=.75\textwidth]{stagnation_vs_solution.png}
%\caption{Various scenarios for trivial flow.  The red line corresponds to the vorticity in the upper fluid, while the yellow line corresponds to the linear shear in the lower fluid.  Lines are extended to illustrate when stagnation enters the fluid domain through the interface vs. through the rigid lids.}\label{fig:variousConfigs}
%\end{figure}

%-------------------------------------------------------------------------------------------------------------
%-------------------------------------------------------------------------------------------------------------
\subsection{Constructing Solutions: Numerical Implementation}
%-------------------------------------------------------------------------------------------------------------
%-------------------------------------------------------------------------------------------------------------

	Using the above formulation, we numerically determine traveling wave solutions by solving for the set $(\eta,\partial_x q_2, c)$ for solutions with a fixed spatial period $L$ and for fixed wave-height $\|\eta\|_\infty$.  In the following section, we outline the specific details of the numerical implementation of our pseudo-arclength continuation method\footnote{Code is available on GitHub.}.

	To numerically solve \eqref{eqn:nonlocalTWEA} \& \eqref{eqn:nonlocalTWEB}, we use a pseudospectral method to solve for the Fourier coefficients of our unknown functions $\eta$ and $\partial_xq_2$ -- differentiation is conducted in Fourier space, while nonlinear operations are computed in physical space.  Since we are considering $2\pi$ periodic solutions, we numerically represent $\eta$ and $\partial_x q_2$ by their Fourier series representation with $N$ Fourier modes of the form $$\eta(x) = \sideset{}{'}\sum_{n = -N}^N\hat{\eta}_n e^{ik_n x}, \qquad \partial_xq_2(x) = \sideset{}{'}\sum_{n = -N}^N \hat{Q}_n e^{ik_n x},$$  where the $^\prime$ denotes the summation with $n\neq 0$ as we have both eliminated the average value of $\eta$ as well as the average value of $\partial_xq_2$.  Enforcing $\partial_xq_2$ has zero average, we consider zero mean-flow in both the upper and lower fluid.  Thus, both $\eta$ and $\partial_x q_2$ are represented by $2N$ unknown Fourier coefficients.  

	Evaluating Equations \eqref{eqn:nonlocalTWEA} and \eqref{eqn:nonlocalTWEB} for $k = k_{-N},\ldots, k_{-2},k_{-1},k_1,k_2,\ldots k_N$ generates $2\times(2N)$ equations for the $2\times(2N)$ unknowns $\hat\eta_n$ and $\hat{Q}_n$ for $n = -N,\ldots,-1,1,\ldots,N$.  In order to solve for traveling waves solutions with various amplitudes, we enforce a fixed amplitude for the interface $\eta$ by allowing the wave-speed $c$ to vary as a function of the amplitude, thereby introducing both a new equation and new unknown into our system.

	Equations (\ref{eqn:nonlocalTWEA}-\ref{eqn:nonlocalTWEB}) are solved for $(\hat{\eta},\hat{Q},c)$ iteratively via Newton's method (though other iterative techniques can also be used).  Using a pseudo-arclength continuation, we determine traveling wave solutions for increasingly larger amplitudes.

	\begin{rmk} Due to the exponential nature of the hyperbolic sine and cosine functions, we rewrite the quantities $\mathcal{C}_j$ and $\mathcal{S}_j$ for $j = 1, 2$ as follows
		$$\mathcal{C}_1 = \cosh(\mu k (\epsilon\eta-1)) = \cosh(\mu k)\left(\cosh(\mu k \epsilon\eta) - \tanh(\mu k)\sinh(\mu k \epsilon\eta)\right)$$
		$$\mathcal{S}_1 = \sinh(\mu k (\epsilon\eta-1)) = \cosh(\mu k)\left(\sinh(\mu k \epsilon\eta) - \tanh(\mu k)\cosh(\mu k \epsilon\eta)\right)$$
		\end{rmk}

\subsubsection{Visualizing Streamlines and Pressure Contours}
For periodic traveling wave solutions, the solutions for the streamlines take the form $$\psi_j = a_0 + a_1(z + (-1)^j\delta_j) -\frac{\omega_j}{2}\left(z + (-1)^j\delta_j\right)^2 + \sum_{k=-\infty}^{\infty}e^{ikx}\hat\psi^{(j)}_k\sinh(k(z+(-1)^j\delta_j)).$$  Choosing the nondimensional function $$\varphi_j = e^{-ikx}\cosh(\mu k(z + (-1)^j\delta_j)),$$ where $\delta_1 = 1$, and $\delta_2 = \delta$, then the analogue of \eqref{eqn:nonlocalInt} for stationary solutions in a traveling frame is, for $k\neq0$, given by 
\begin{align}
	&\int_0^{2\pi}e^{-i k x}\left(\cosh(\mu k(\epsilon\eta + (-1)^j\delta_j))\left(\partial_xq_j - \omega_j \eta - c\right) + \frac{\omega}{k}\sinh(\mu k (z + (-1)^j\delta_j)\right)\:dx\nonumber\\
	&\qquad = \int_0^{2\pi}e^{-i k x} \partial_z\psi_j(x,(-1)^j\delta_j)\:dx
	\end{align}
whereas for $k = 0$, we find 
\begin{equation}
	\int_0^{2\pi}\partial_z\psi_j(x,(-1)^j\delta_j)\:dx =\int_0^{2\pi}e^{-i k x}\left(\partial_xq_j - c \right)\:dx.
	\end{equation}
Once we have determined a TWS corresponding to the set $(\eta, q_1, q_2, c)$, we can then use the above equations to determine the streamfunction within the bulk of each fluid by solving for the coefficient $a_1$ as well as $\hat\psi_k^{(j)}$.  We also note that the streamfunction $\psi$ can only be determined up to an arbitrary constant within the fluid domain.  Here we choose to normalize so that the streamfunctions are zero on the interface.  This flexibility allows us to determine the value of $a_0$.

\subsection{The Irrotational Case}
\begin{figure}
\includegraphics[width=.45\textwidth]{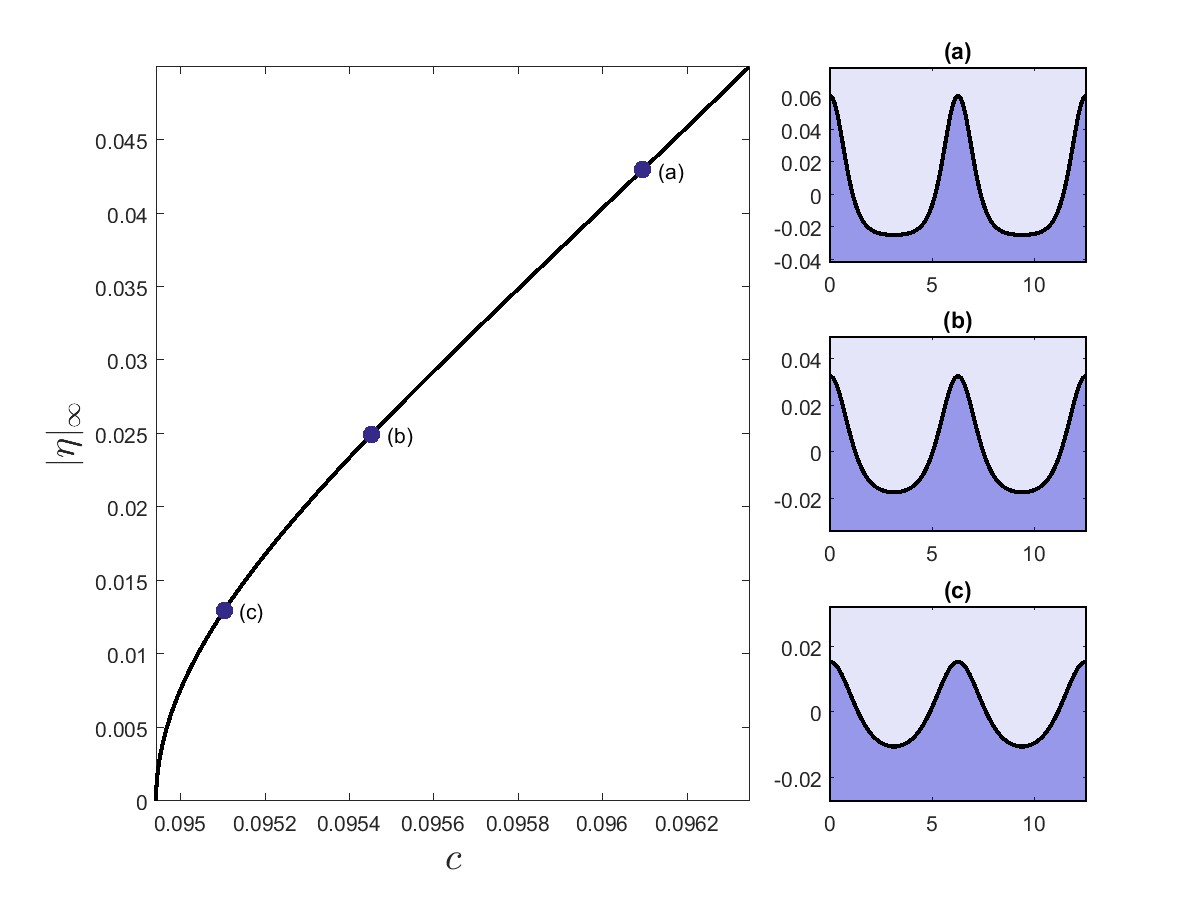}
\includegraphics[width=.45\textwidth]{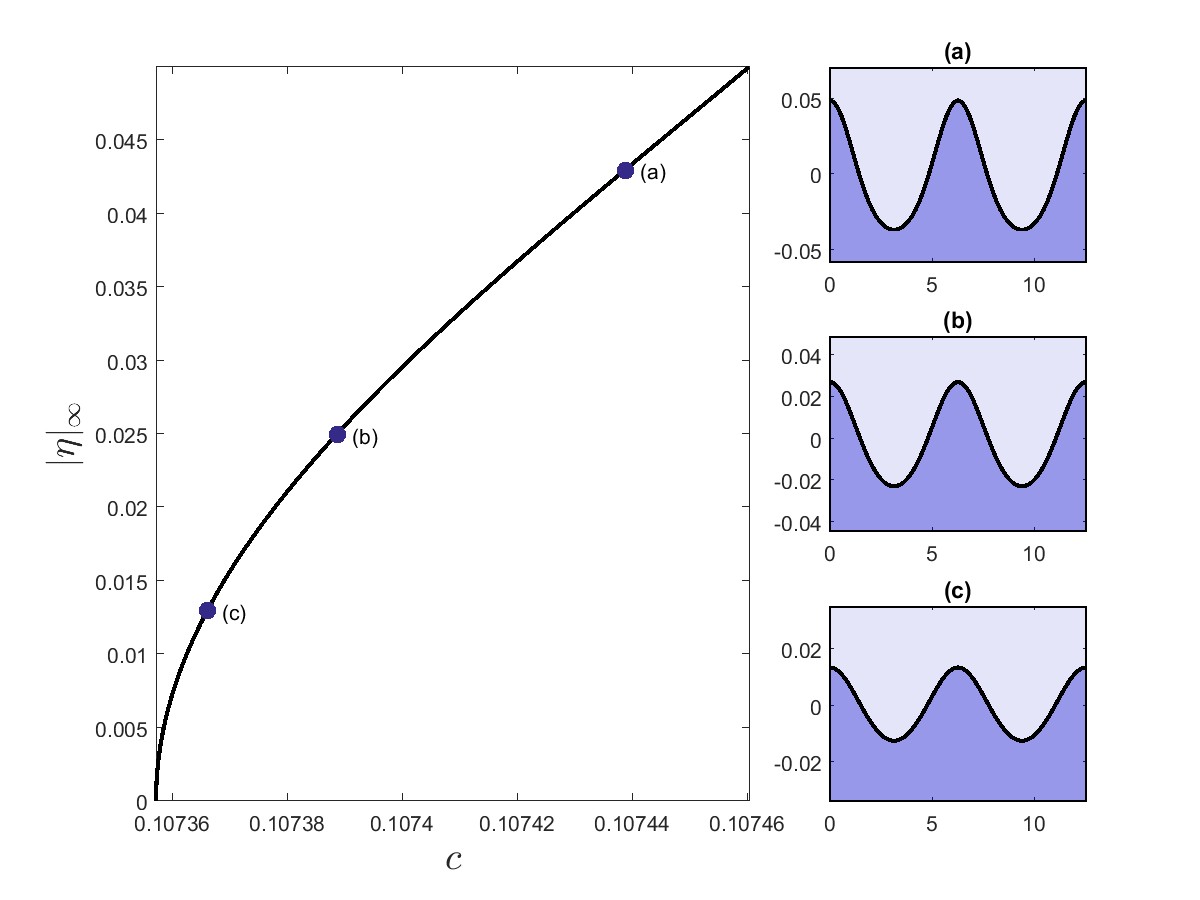}\\
\includegraphics[width=.45\textwidth]{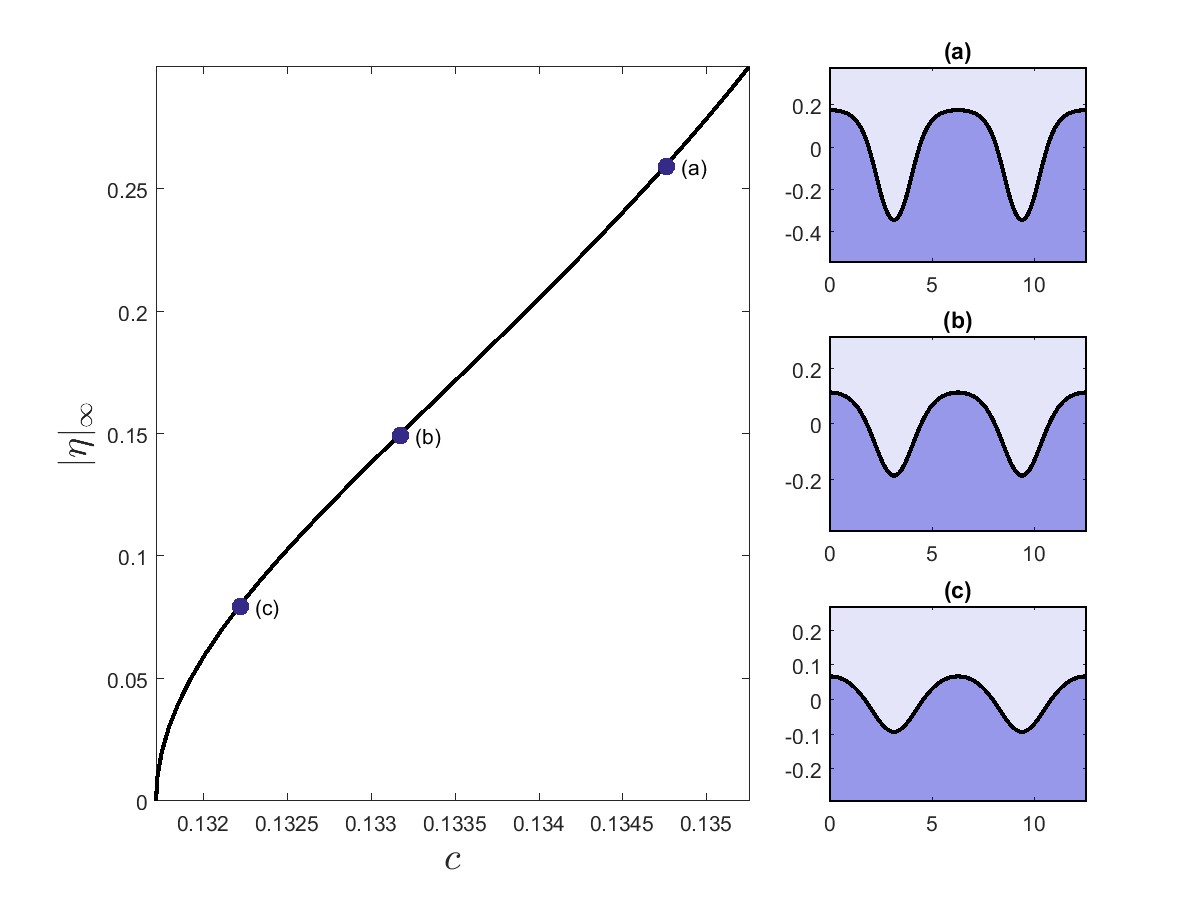}
\includegraphics[width=.45\textwidth]{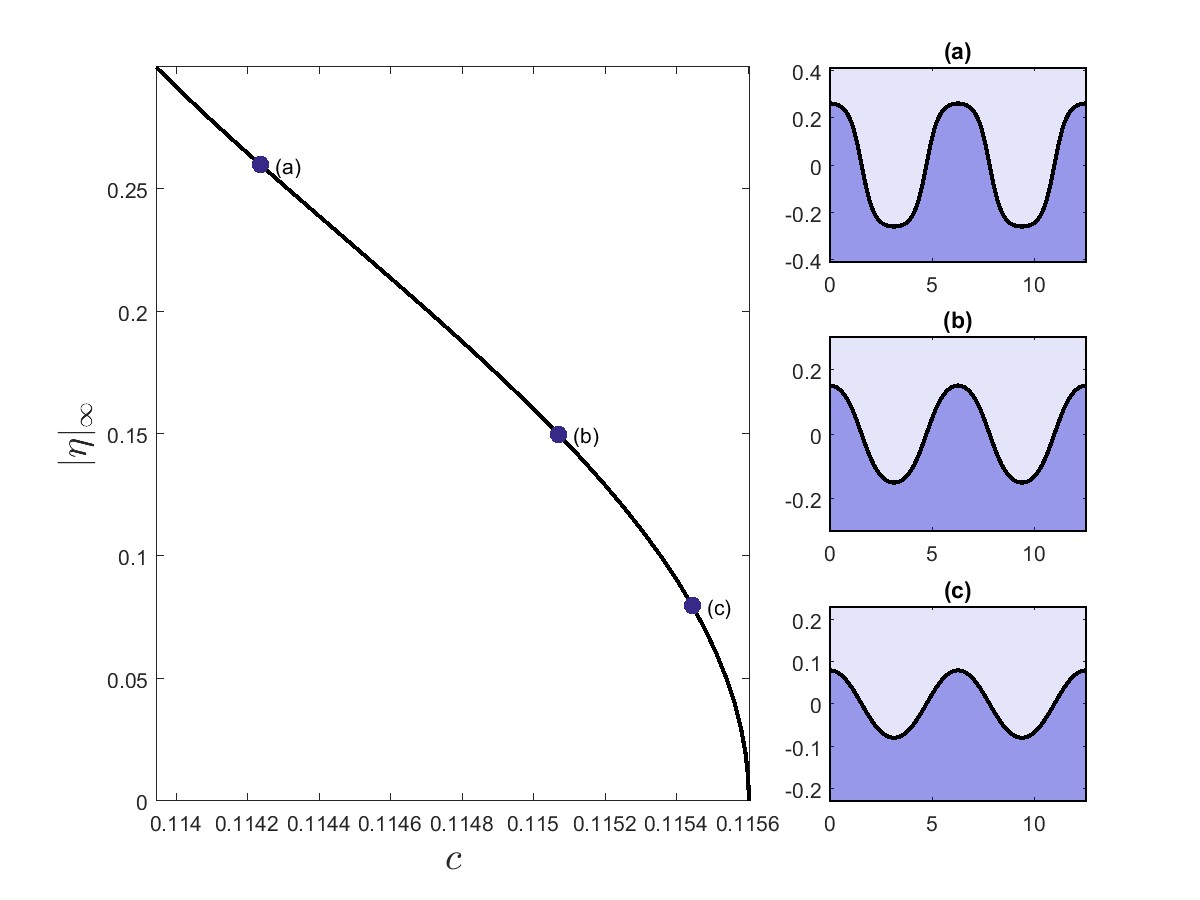}
\caption{Bifurcation curves and solutions for $\omega_1 = \omega_2 = 0$, $\mu = \sqrt{.1}$, and $\rho = 1.028$ and $\delta = .5$, $\delta = .75$, $\delta = 1$ and $\delta = 2$ starting clockwise from the top.}\label{fig:irrotational_sols_lowrho}
\end{figure}

Before we investigate the effects of shear in each layer, we begin by investigating solutions with no shear (vorticity) in each layer. In Figure \ref{fig:irrotational_sols_lowrho}, we examine the weak stratifications for $\rho = 1.028$, $\mu = \sqrt{0.1}$ and various values of the depth-ratio $\delta$.  For $\delta = 0.5$, as we increase the amplitude, both the speed of the traveling wave, and the interface becomes more peaked.  However, as $\delta$ increases from $\delta = 0.5$ to $\delta = 2$, both the amplitude-speed dependence and qualitative shape of the solution undergo a transformation.  

	Specifically, for $\delta = 0.5$, waves with larger amplitude travel at higher speeds.  However, as $\delta$ increases, this relationship changes so that waves of larger amplitude travel at sower speeds than those with lower amplitudes.  As $\delta$ continues to increase, the amplitude-speed relationship continues to change until the waves with larger amplitudes again travel at larger speeds.  Simultaneously, the qualitative shape of solutions morph from ``peaked waves of elevation'', through rounded peaks and troughs, and finally to ``peaked waves of depression''.  This transition can be accurately predicted by examining how the sign of $c_2$ varies with respect to $\mu$ and $\delta$ for a specific value of $\rho$ as seen in Figure \ref{fig:turningPtMuvsDel}.

\subsection{Including Shear Effects}
We now consider the influence of linear shear within each fluid layer.  In particular, we focus on the difference between the shape of solutions and the overall bifurcation structure of traveling wave solutions.  Specifically, we examine solutions for $\delta = 4$, $\rho = 1.028$ and $\mu = \sqrt{.1}$ that bifurcate from trivial flows that various configurations of stagnation points within each layer of the fluid.  

Noting that we numerically solve for both the free surface $\eta$ and the tangential velocity $q_2(x)$, we can conjecture a limiting wave-height by imposing that $q_1(x)$
%%%%%%%%%%%%%%%%%%%%%%%%%%%%%%%%%%%%%%%%
\subsubsection{Bifurcating from trivial flow with no stagnation points in either fluid.}
\begin{figure}
\centering
% \begin{tabular}{c}
% \includegraphics[width=.45\textwidth]{solutions_epsilon_1_mu_0_31623_delta_4_w1_1_w2_m_1_rho_1_028_ampStart_0_0001_ampEnd_0_01.jpg}\\
% (a)\\
% \includegraphics[width=.75\textwidth]{streamlines_epsilon_1_mu_0_31623_delta_4_w1_1_w2_m_1_rho_1_028_ampIter_40_halfMaxMin_0_0038704.png}
% \\
% (b)
% \end{tabular}
\hspace*{-.5in}\includegraphics[width=1.2\textwidth]{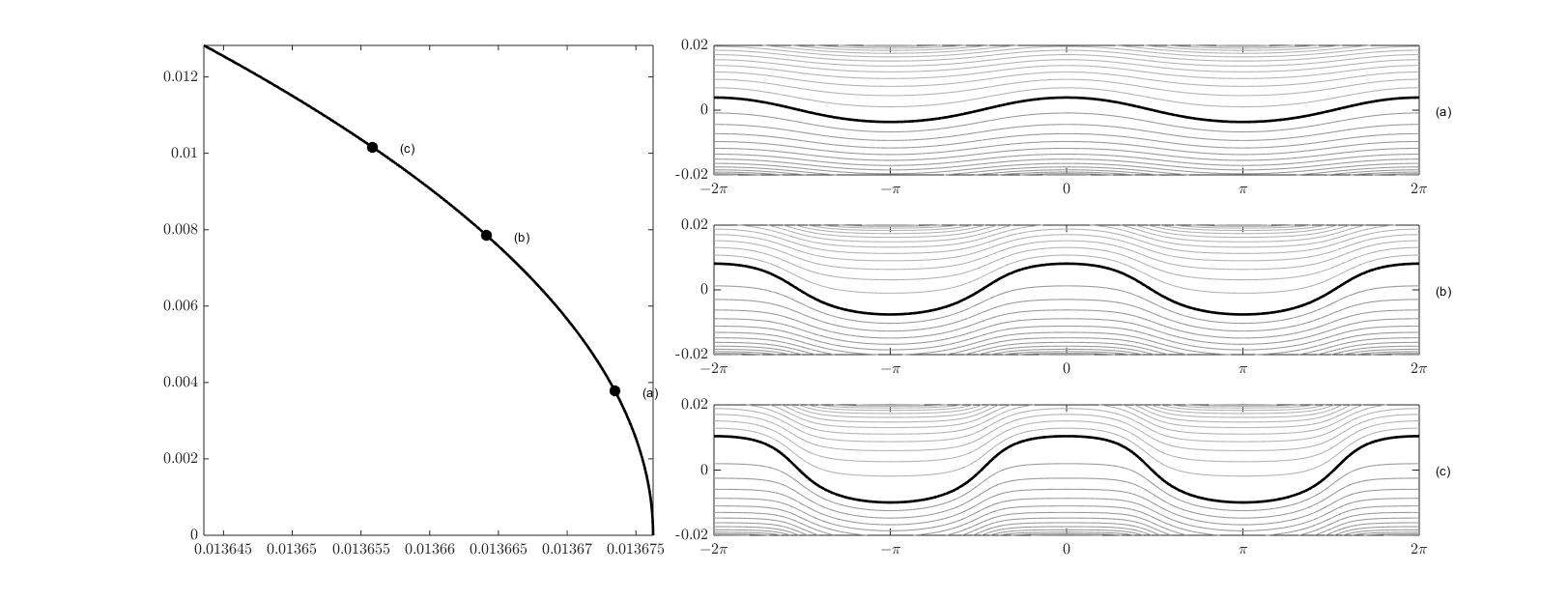}
\caption{$\mu = \sqrt{.1}$, $\rho = 1.028$, $\delta = 4$, $\omega_1 = 1$, and $\omega_2 = -1$, with amplitude-speed relationship in (a) and streamlines corresponding to wave amplitudes of  $\eta\approx 0.0039$ (b).}
\label{fig:nostagpts}
\end{figure}

To look at an example in which we bifurcate from a flat surface with a stagnation point in the lower fluid, we choose $\omega_{1}=1$ and $\omega_{2}=-1$.  As seen in Figures \ref{fig:nostagpts} (a) and (b), the abscence of stagnation points does not appear to bias the interface towards elevation or depression, thus allowing for the development of far more symmetric nonlinear wave profiles than are seen in subsequent sections.  We also note that, by referring to Figure \ref{fig:lowrho}, the choice of vorticities corresponds to a negative value of $c_{2}$.  This is corroborated by the speed-amplitude curve in Figure \ref{fig:nostagpts} (a), which shows the speed of the wave decreasing with the increase in the amplitude.  

%%%%%%%%%%%%%%%%%%%%%%%%%%%%%%%%%%%%%%%%
\subsubsection{Bifurcating from trivial flow with a stagnation point in the upper fluid.}
\begin{figure}
\centering
% \begin{tabular}{c}
% \includegraphics[width=.45\textwidth]{solutions_epsilon_1_mu_0_31623_delta_4_w1_m_1_w2_m_1_rho_1_028_ampStart_0_001_ampEnd_0_0098371.jpg}\\
% (a)\\
% \includegraphics[width=.75\textwidth]{streamlines_epsilon_1_mu_0_31623_delta_4_w1_m_1_w2_m_1_rho_1_028_ampIter_40_halfMaxMin_0_026099.png}
% \\
% (b)
% \end{tabular}
\hspace*{-.5in}\includegraphics[width=1.2\textwidth]{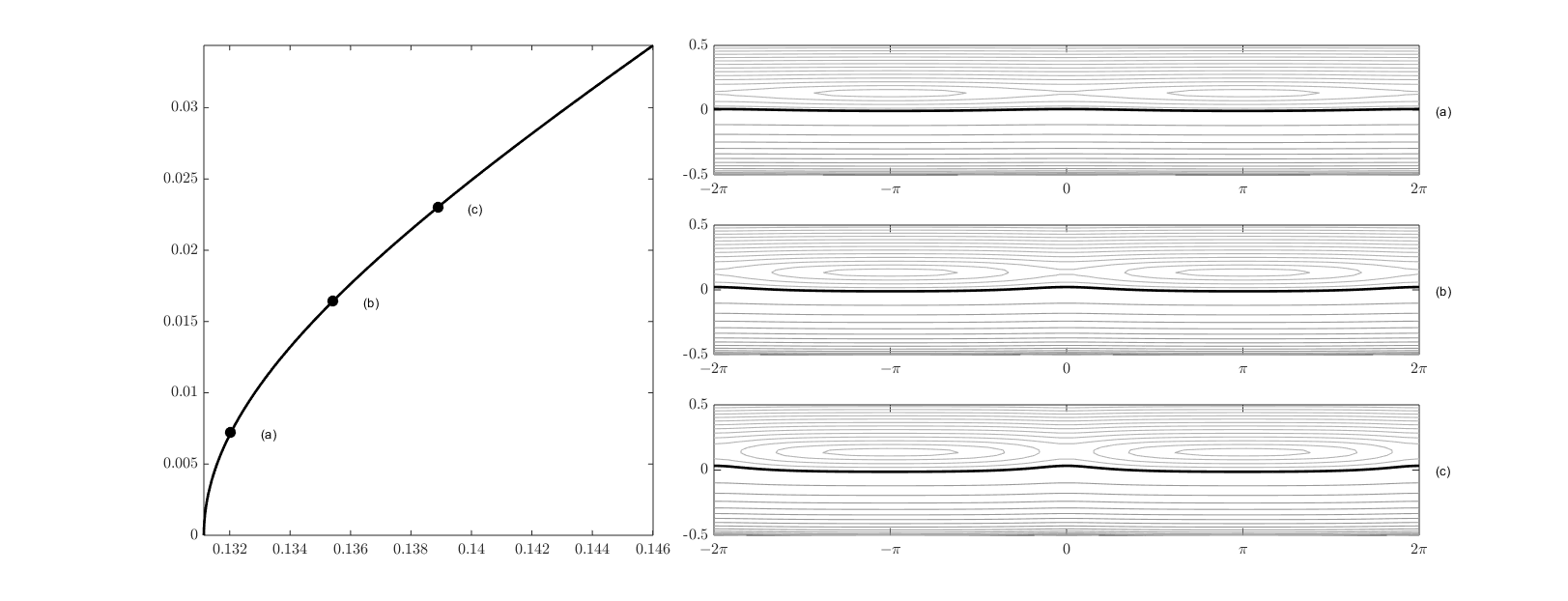}
\caption{$\mu = \sqrt{.1}$, $\rho = 1.028$, $\delta = 4$, $\omega_1 = -1$, and $\omega_2 = -1$, with amplitude-speed relationship in (a) and streamlines corresponding to wave amplitudes of  $\eta\approx 0.0261$ (b).  The stagnation point in the upper fluid region is seen at the saddle centered around $x=0$ as well as at the centers located at $x = \pm \pi$.}
\label{fig:stagup}
\end{figure}

To look at an example in which we bifurcate from a flat surface with a stagnation point in the lower fluid, we choose $\omega_{1}=\omega_{2}=-1$.  As can be seen by examing the interface profiles in Figure \ref{fig:stagup} (a), the particular case we have chosen corresponds to interfaces of elevation.  By examing the impact of the stagnation point on the streamline patterns seen in Figure \ref{fig:stagup} (b), we can see from where this tendency towards forming interfaces of elevation comes.

%%%%%%%%%%%%%%%%%%%%%%%%%%%%%%%%%%%%%%%%
\subsubsection{Bifurcating from trivial flow with a stagnation point in the bottom fluid.}
\begin{figure}
\centering
% \begin{tabular}{c}
% \includegraphics[width=.45\textwidth]{solutions_epsilon_1_mu_0_31623_delta_4_w1_1_w2_1_rho_1_028_ampStart_0_0001_ampEnd_0_0097533}
% \\
% (a)\\
% \includegraphics[width=.75\textwidth]{streamlines_epsilon_1_mu_0_31623_delta_4_w1_1_w2_1_rho_1_028_ampIter_30_halfMaxMin_0_035369.png}
% \\
% (b)
% \end{tabular}
\hspace*{-.5in}\includegraphics[width=1.2\textwidth]{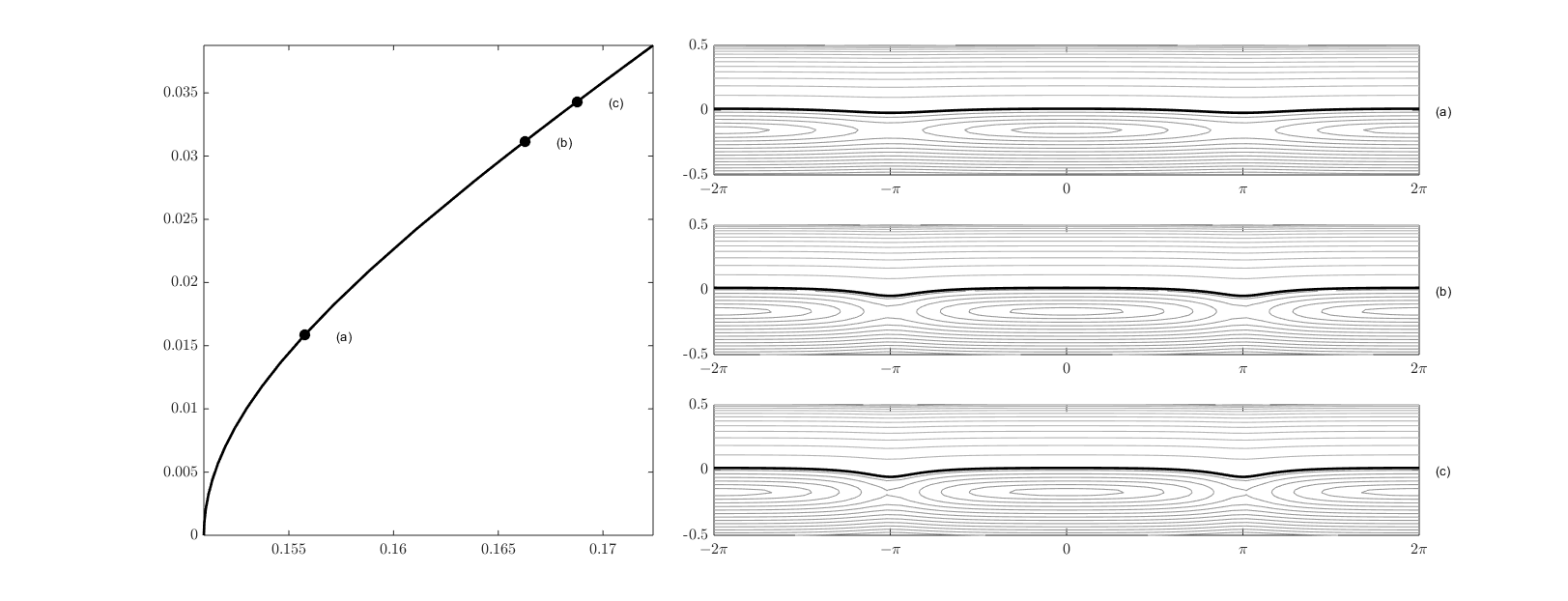}
\caption{$\mu = \sqrt{.1}$, $\rho = 1.028$, $\delta = 4$, $\omega_1 = 1$, and $\omega_2 = 1$, with amplitude-speed relationship in (a) and streamlines corresponding to a wave amplitude of  $\eta\approx 0.0354$ (b). The stagnation point in the lower fluid region is seen at the saddles centered around $x=\pm \pi$ and the center located at $x = 0$.}
\label{fig:staglow}
\end{figure}

To look at an example in which we bifurcate from a flat surface with a stagnation point in the lower fluid, we choose $\omega_{1}=\omega_{2}=1$.  As can be seen by examing the interface profiles in Figure \ref{fig:staglow} (a), the particular case we have chosen corresponds to interfaces of depression.  By examing the impact of the stagnation point on the streamline patterns seen in Figure \ref{fig:staglow} (b), we can see from where this tendency towards forming interfaces of depression comes.

%%%%%%%%%%%%%%%%%%%%%%%%%%%%%%%%%%%%%%%%
\subsubsection{Bifurcating from trivial flow with a stagnation point in both fluids.}
\begin{figure}
\centering
\hspace*{-.5in}\includegraphics[width=1.2\textwidth]{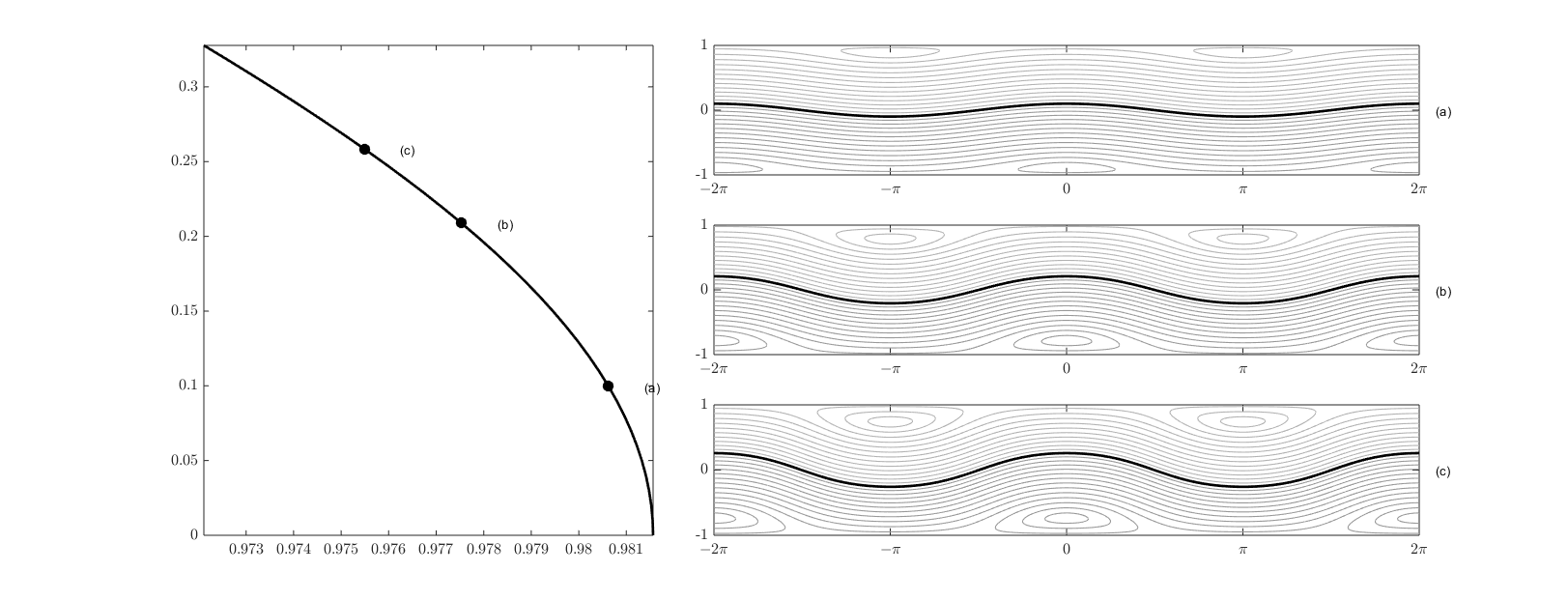}
\caption{$\delta = 1$, $\omega_1 = -1$, $\omega_2 = 1$, $\rho = 1.028$,  showing both stagnation points in the upper and lower fluid centered at $x = 0$ in the lower fluid, and $x = \pm\pi$ in the upper fluid.}
\label{fig:twostagpts}
\end{figure}

To see an example in which we bifurcate with stagnation points in both the lower and upper fluid regions, we choose $\omega_{1}=-1$ and $\omega_{2}=1$.  In this case, we see that the tendency is for waves of depression to form which increase in speed with amplitude; see Figure \ref{fig:twostagpts} (a).  The role of stagnation becomes increasingly more relevant on the nonlinear profile as its amplitude is increased, and thus has an ever increasing effect on the shape of the wave profile; compare Figures \ref{fig:twostagpts} (b) and (c).  For lower amplitudes, the presence of stagnation points both above and below the interface allows for the kind of ambiguity between the interface being elevated or depressed for which the case of no stagnation points in the fluid allowed.  However, the presence of stagnation points allows for the formation of broader rising portions of heavier fluid, which at larger amplitudes arguably makes the interface one of depression with sharper falling peaks that are shaped by having to move between the stagnation points in the fluid. 

% = = = = = = = = = = = = = = = = = = = = = = = = = = = = = = = = = = = = = = = = = = = = = = = = = = = = = = = = = = 
\section{Stability Formulation}\label{sec:stability}
% = = = = = = = = = = = = = = = = = = = = = = = = = = = = = = = = = = = = = = = = = = = = = = = = = = = = = = = = = = 

	We investigate the spectral stability of the traveling wave profiles computed in the previous section with respect to infinitesimal perturbations.  First, it is important that we discuss what perturbations we wish to allow. It may appear natural to consider disturbances of the same period as the underlying stationary wave, as is often done in the literature.  However, we wish to work with a more general class of disturbances; namely, we choose to include \textit{all} bounded on the whole real line.  It is important to realize that this class is the largest class of perturbations allowed by the physical problem at hand. Indeed, disturbances should be bounded and continuous functions, but there is no physical reason to restrict their spatial dependence to be periodic.

	\subsection{Stability Formulation: Problem Reformulation}
	In order to investigate the stability of the traveling wave profiles with respect to such perturbations, it is necessary to reformulate the governing equations. Equation \eqref{eqn:bernoulliNonDim} is a local statement and does not require modification. However, \eqref{eqn:TopNonlocalNonDim} and \eqref{eqn:BottomNonlocalNonDim} are nonlocal and in their current incarnation, apply specifically to waves of period $L$. Thus, we cannot use the equation in its current form for any bounded function on the real line.

	Following the method outlined in \cite{oliveras}, we reformulate the nonlocal equations on the whole-line via a spatial average value.  For a continuous bounded function $f(x)$, i.e. $f \in C^{0}_{b}(\mathbb{R})$, let $\left<f\right>$ represent the spatial average of the function defined by
	\[
	\left<f\right>=\lim_{M\rightarrow \infty} \frac{1}{M}\int_{-M/2}^{M/2} f(x)\: dx.
	\]

	\noindent It should be noted that the kernel of this operation is quite large. For instance, all functions that approach zero as $|x|\rightarrow \infty$ have zero spatial average. Nevertheless, we may use the spatial average to replace \eqref{eqn:TopNonlocalNonDim} and \eqref{eqn:BottomNonlocalNonDim} with the more general nonlocal equations given by

		\begin{equation}{}
			\left<e^{-ikx}\left(\left(ik\epsilon\mu^2\partial_t\eta - \omega_1 \right)\mathcal{C}_1-\mu k\left(\epsilon(\partial_xq_1 - \omega_1\eta)-c\right)\mathcal{S}_1\right)\right> = 0,\label{eqn:nonlocal_top_spatial}
			\end{equation}
	and

			\begin{equation}
				\left<e^{-ikx}\left(\left(ik\epsilon\mu^2\partial_t\eta - \omega_2 \right)\mathcal{C}_2-\mu k\left(\epsilon(\partial_xq_2 - \omega_2\eta)-c\right)\mathcal{S}_2\right)\right> = 0,\label{eqn:nonlocal_bottom_spatial}
				\end{equation}

	\noindent where $\mathcal{S}_j$ and $\mathcal{C}_j$ are defined as before.  These equations are valid for all $k\in \mathbb{R}_0=\mathbb{R}-\{0\}$, as no quantitization condition is imposed by the periodicity of the solutions considered. Further, as solutions of increasingly larger period are considered, the set of $k$ values to be considered in \eqref{eqn:nonlocal_top_spatial} and \eqref{eqn:nonlocal_bottom_spatial} approaches a dense subset of the real line. The equation corresponding to $k=0$ is excluded, as before.  Equations \eqref{eqn:nonlocal_top_spatial} and \eqref{eqn:nonlocal_bottom_spatial} allows us to perturb our traveling wave solution with any bounded perturbation regardless of periodicity.

%-------------------------------------------------------------------------------------------------------------
%-------------------------------------------------------------------------------------------------------------
\subsection{Stability Formulation: Eigenvalue Problem}
%-------------------------------------------------------------------------------------------------------------
%-------------------------------------------------------------------------------------------------------------
	Having generalized the dynamical equations to accommodate the perturbations we wish to consider, we briefly discuss the definition of spectral stability.  A stationary solution of a nonlinear problem is spectrally stable if there are no exponentially growing modes of the corresponding linearized problem.  To determine the spectral stability of the periodic traveling wave solutions, we start by considering a traveling wave solution set $(\eta_0(x-ct)$, $\alpha_0(x-ct)$,$\beta_0(x-ct)$). In the same traveling coordinate frame, we add a small perturbation of the form
	\begin{eqnarray*}
	\eta(x-ct,t) &=& \eta_0(x-ct) + \varepsilon \eta_1(x-ct)e^{\lambda t}+ \mathcal{O}(\varepsilon^2),\\
	q_1(x-ct,t)  &=& \alpha_0(x-ct) + \varepsilon \alpha_1(x-ct)e^{\lambda t} + \mathcal{O}(\varepsilon^2),\\
	q_2(x-ct,t)  &=& \beta_0(x-ct) + \varepsilon \beta_1(x-ct)e^{\lambda t} + \mathcal{O}(\varepsilon^2),
	\end{eqnarray*}

	\noindent where $\varepsilon$ is a small parameter.  The perturbations $\eta_1$, $\alpha_1$, and  $\beta_1$ are moving at the same speed and in the same direction as the original traveling wave solution.  Our goal is to determine the time dependence of the perturbation in order to determine how the deviation from the unperturbed solution evolves.

%-------------------------------------------------------------------------------------------------------------
%-------------------------------------------------------------------------------------------------------------

As we are linearlizing about a traveling wave solution, we substitute the above expansions for $q_1$, $q_2$ and $\eta$ into \eqref{eqn:bernoulliNonDim}, \eqref{eqn:nonlocal_top_spatial}, and \eqref{eqn:nonlocal_bottom_spatial} keeping only $\mathcal{O}(\varepsilon)$ terms. This is rewritten compactly as

\begin{equation}\label{eqn:sovc}
\lambda \mathcal{L}_1U(x) = \mathcal{L}_2U(x),
\end{equation}

\noindent where $\mathcal{L}_{1}$ and $\mathcal{L}_{2}$ are $3\times 3$ matrices of linear operators and $U(x)$ is a vector-valued function with entries $U(x) = \left[\eta_1~~\alpha_1~~\beta_1\right]^{T}$.  Details are provided in Appendix A.  

Since the time dependence of the perturbation depends exponentially on $\lambda$, we can determine information about the stability of the underlying traveling wave by determining all bounded solutions of this generalized eigenvalue problem.  If any bounded solutions $U(x)$ exist for which the corresponding $\lambda$ has a positive real part, the linear approximation of the solution will grow in time and thus the perturbed solution will exponentially diverge from the stationary solution in the linear approximation.

Since the coefficient functions of (\ref{eqn:sovc}) are periodic in $x$ with period $2\pi$, we decompose the perturbations further using Floquet's Theorem, see for instance \cite{decon,curtis2}.  This allows us to further decompose $\eta_1$, $\alpha_1$, and $\beta_1$ in the form $$\eta_1(x) = e^{i p x}\bar{\eta}_1(x), \qquad \alpha_1(x) = e^{i p x}\bar{\alpha}_1(x), \qquad \beta_1(x) = e^{i p x}\bar{\beta}_1(x)$$
\noindent where $\bar{\eta}_1(x)$, $\bar{\alpha}_1$, and $\bar{\beta}_1$ are periodic with period $2\pi$ and $p$ (the Floquet exponent) can be restricted to the interval $[-1/2,1/2]$.

Substituting the Floquet decomposition directly into \eqref{eqn:sovc} while simultaneously using the Fourier series representation for $\bar\eta_0$, $\bar\alpha_0$ and $\bar\beta_0$ as outlined in \cite{oliveras},  we obtain a new equation of the form 
\begin{equation}
	\lambda \tilde{\mathcal{L}}_1(p)\hat{U} = \tilde{\mathcal{L}}_2(p)\hat{U}, \label{eqn:LthreeForm}
	\end{equation}
where $\hat{U}$ represents the concatenation of three bi-infinite vectors of the Fourier coefficients for $\bar\eta_1$, $\bar\alpha_1$ and $\bar\beta_1$.% under the assumption that $\bar\eta_1$, $\bar\alpha_1$, and $\bar\beta_1$ are zero-average value functions.  
For $j = 1, 2$, $\tilde{\mathcal{L}}_j(p)$ are linear operators that now depend on the Floquet exponent $p$. Equation \eqref{eqn:LthreeForm} gives a generalized bi-infinite eigenvalue problem for determining the spectrum of the linearized operator about the stationary traveling wave solutions.  

Instead of directly solving the eigenvalue problem as stated above, a more stable approach is to reformulate the problem as a quadratic eigenvalue problem for the eigenvalue $\lambda$ and the corresponding eigenfunction $\bar\eta_1$ of the form
\begin{equation}\left(\lambda^2\mathcal{A}_2(p) + \lambda\mathcal{A}_1(p)+\mathcal{A}_0(p)\right)\bar\eta_1 = 0.\label{eqn:polyEig}
\end{equation} Details regarding \eqref{eqn:polyEig}, the form of the $\mathcal{A}_j$'s for $j = 0, 1,$ and $2$ are given in Appendix B.

We solve this generalized eigenvalue problem numerically by truncating the Fourier series representation for $\bar\eta_1$ over $\mathbb{Z}-\lbrace 0\rbrace$ to $\left\{\pm 1, \pm 2, \ldots, \pm N \right\}$ where we have eliminated the zero-mode as we only consider zero-average perturbations to the free-surface.  With this truncation, we obtain $2N$ quadratic equations for $\cdot2N$ unknown Fourier coefficients of $\bar\eta_1$.   Finally, we note that due to the underlying symmetries in the problem, we may restrict the $p$ interval from $\left[-1/2, 1/2\right]$ to $\left[0,1/2\right]$ and consider both $\lambda$ and the conjugate of $\lambda$.  Thus, for every $p\in\left[0,1/2\right]$, we solve the generalized eigenvalue problem given by 
\begin{equation}
	\left(\lambda^2\mathcal{A}^{(N)}_2(p) + \lambda\mathcal{A}^{(N)}_1(p)+\mathcal{A}^{(N)}_0(p)\right)\hat{\bar\eta}_1^{(N)} = 0. \label{eqn:LthreeFormTrunc}
	\end{equation}
where the superscript $(N)$ denotes the projection on to the $N$ Fourier modes. This is done via the QZ algorithm \cite{moler}.  The dimension of the truncation on the Fourier modes is chosen so that both the eigenvalues and eigenvectors converge to 12 digits of accuracy.

% = = = = = = = = = = = = = = = = = = = = = = = = = = = = = = = = = = = = = = = = = = = = = = = = 
\section{Stability Results}
% = = = = = = = = = = = = = = = = = = = = = = = = = = = = = = = = = = = = = = = = = = = = = = = = 
% !tex root = ../rigid_lid_stability_v1.tex

Using the formulation outlined in the previous section, we numerically compute spectra beginning with irrotational interfacial waves.  It is well known that the trivial solution is subject to the Kelvin-Helmholtz instability provided that there is a jump in the horizontal velocity at the interface. However, the role that nonlinearities play in the formation of possible Kelvin-Helmholtz instabilities in our configuration is not as well known.  In what follows then, we examine the role nonlinearity plays in the formation of instabilities by examining the spectra associated with the non-trivial TWSs presented in the previous section.  We begin by considering the spectral stability corresponding to the irrotational problem and then subsequently explore the impact of linear shear.

It is straight-forward to show that the trivial solution will be spectrally stable when there is no jump in the tangential velocity at the interface.  However, the spectral stability is not guaranteed for non-trivial solutions that bifurcate from this solution. Throughout this section, we consider two separate classes of instabilities: modulational instabilities (MIs) corresponding to long-wave perturbations ($p \sim 0$), and high-frequency instabilities (HFIs) corresponding to the Floquet parameter $p\gg0$.  Note, while computations need only be done for Floquet parameters $p\in[0,1/2]$ due to underlying symmetries in the problem, we plot spectra over $p\in[-1,1]$ for illustrative purposes.    

\begin{figure}
\centering
	\includegraphics[width=\textwidth]{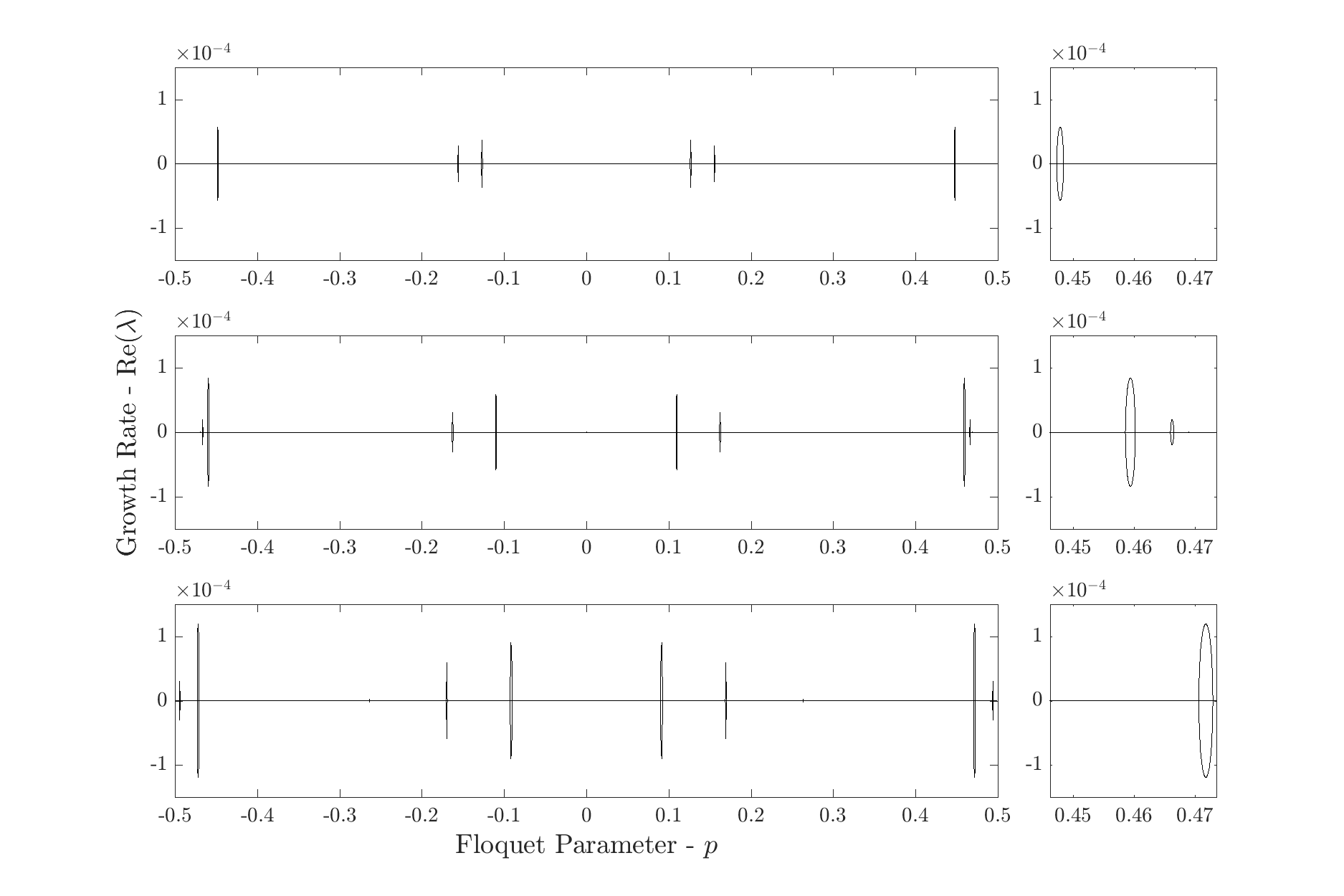}
	\caption{Real part of growth rates as a function of the Floquet parameter $p$ with $\omega_1 = \omega_2 = 0$, $\rho = 1.028$, $\delta = 4$, $\mu = \sqrt{.1}$ and $\epsilon$ increasing from $\epsilon \approx .16$ (top), $\epsilon \approx .18$ (middle), and $\epsilon \approx 0.2$.  The right panel is a zoomed-in version of a region in the left panel and demonstrates that the spikes seen on the left do indeed represent continuous curves in the Floquet parameter.  While narrow bands of HFIs are clearly visible, the absence of spectra near $p\sim 0$ shows MIs are not present.}\label{fig:irrotational_spectra}
	\end{figure}

\begin{figure}
	\centering
	\includegraphics[width=\textwidth]{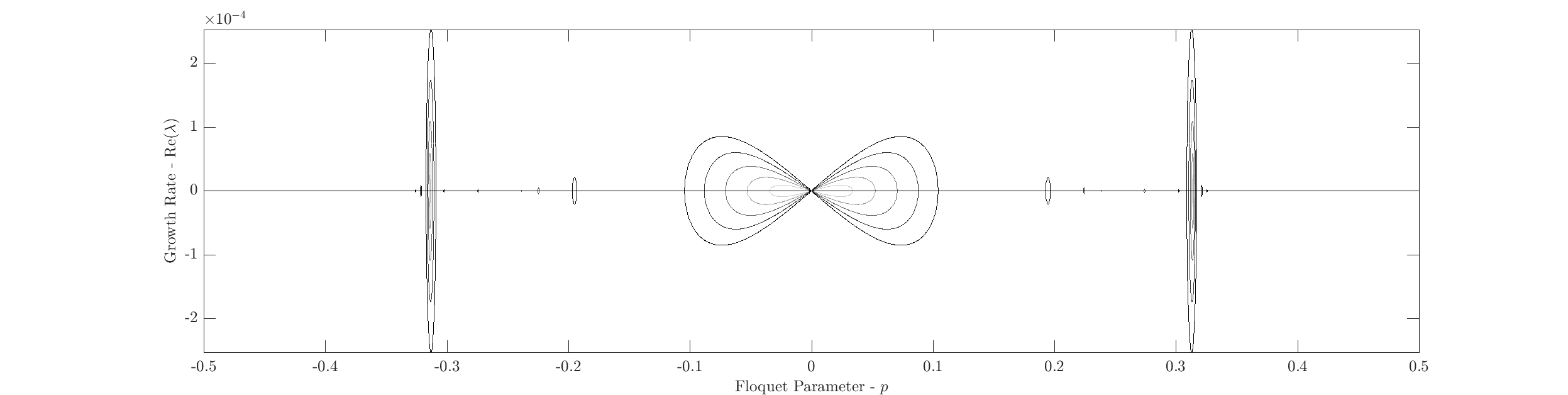}
	\caption{Real part of growth rates as a function of the Floquet parameter $p$ with $\omega_1 =\omega_2 =0$, $\rho = 1.028$, $\delta = 2$, $\mu = 2$ and $\epsilon$ increasing from $\epsilon \approx .001$ (light grey) to $\epsilon \approx .08$ (darker grey). Both MI and HFI are present in this scenario.}\label{fig:irrotational_MI}
	\end{figure}

For the configuration with $\mu = \sqrt{.1}$, $\epsilon = .1$ and $\rho = 1.028$, Figure \ref{fig:irrotational_spectra} shows the maximum growth rate as a function of the Floquet parameter $p$ for increasing values of the amplitude of the traveling wave solution $\eta$.  As in the single-layer free-boundary problem, all non-zero, small-amplitude traveling waves appear unstable to perturbations corresponding to HFIs  localized around small bands of Floquet parameters $p$.  An interesting finding is that for the various depth ratios examined in the irrotational problem with fixed $\mu = \sqrt{0.1}$, all solutions are not susceptible to MIs.  However, if $\mu$ and $\delta$ are varied appropriately, then even the irrotational problem may become susceptible to both HFI and MI.  For example, as seen in Figure \ref{fig:irrotational_MI}, when $\mu = \delta = 2$, $\rho = 1.028$, traveling wave solutions are susceptible to both HFIs and MIs where magnitude of both instabilities increase as a function of increasing $\epsilon$.

\begin{figure}
	\centering
	\includegraphics[width=\textwidth]{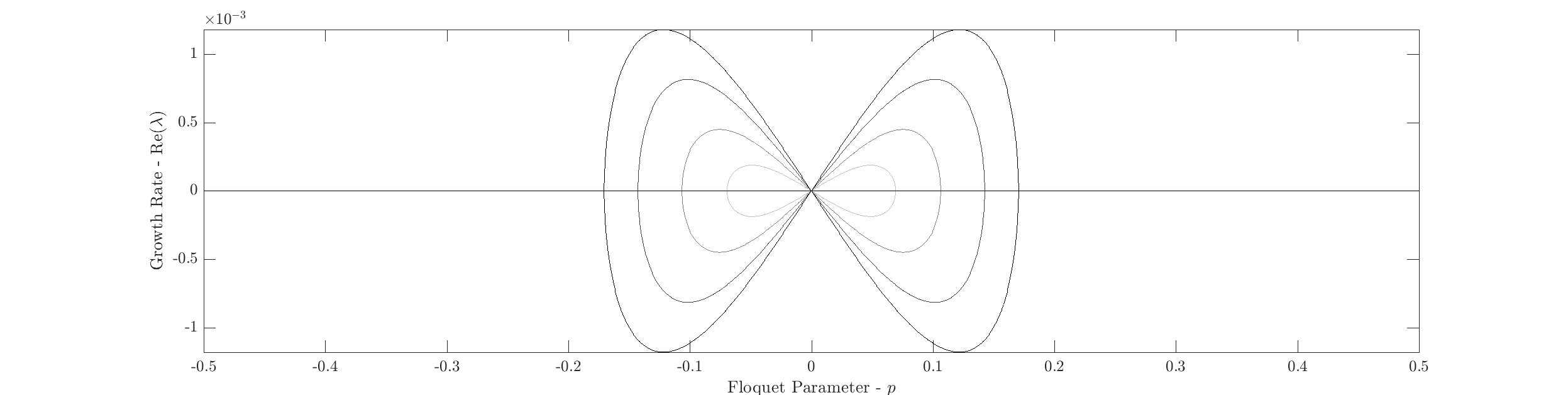}
	\caption{Real part of growth rates as a function of the Floquet parameter $p$ with $\omega_1 =-1$, $ \omega_2 =1$, $\rho = 1.028$, $\delta = 2$, $\mu = 2$ and $\epsilon$ increasing from $\epsilon \approx .001$ to $\epsilon \approx .08$.  While the MI instbilities are present, the magnitude of the HFI have been significantly diminished due to the the presence of the the shear in both layers.}\label{fig:BF_w1_m1_w2_1_delta_1}
	\end{figure}

\begin{figure}
	\centering
	\includegraphics[width=\textwidth]{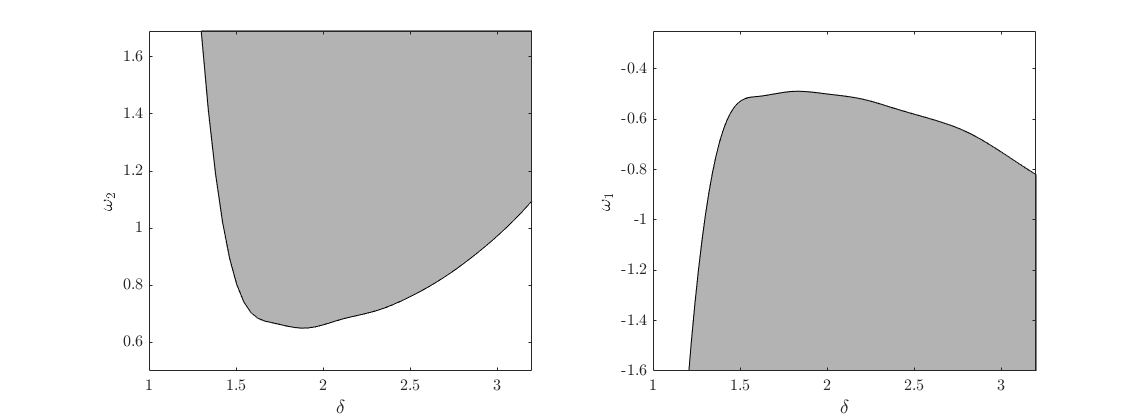}
	\caption{Region of MIs (shaded) as a function of $\delta$ for small amplitude solutions with $\rho = 1.028$ and $\mu = \delta$.  In the left panel, the value of $\omega_1$ is fixed to be zero while the value of $\omega_2$ is varied.  Similarly, in the right panel, the value of $\omega_2$ is fixed to be zero while the value of $\omega_1$ is varied.}
	\label{fig:fixw1w2}
	\end{figure}

\begin{figure}
	\centering
	\includegraphics[width=.75\textwidth]{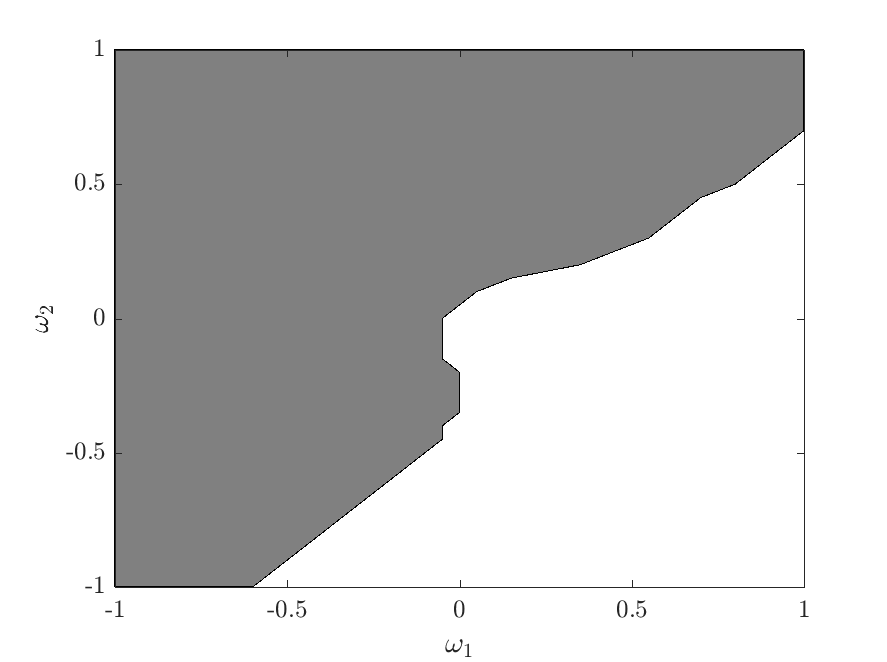}
	\caption{Region of MIs as a function of $\omega_1$ and $\omega_2$ for $\delta = 2$, $\rho = 1.028$, and $\mu = 2$}
	\label{fig:fixdelta}
	\end{figure}

Including vorticity has an impact on the strength of HFI as seen in Figure \ref{fig:BF_w1_m1_w2_1_delta_1}, thereby adding a mechanism for the suppression of instability not seen in the stratified shear-free case.  Furthermore, the inclusion of vorticity may also impact whether or not a traveling wave is susceptible to MI.  This hints at the more complicated role that vorticity plays in understanding and parameterizing regions of instability for the various parameters in the problem.  We explore this issue more fully in Figures \ref{fig:fixw1w2} and \ref{fig:fixdelta}.  Here, we show how different balances between the shear, depth ratio, and $\mu$ either allow or suppress MIs.  

\begin{figure}
\centering
	\includegraphics[width=\textwidth]{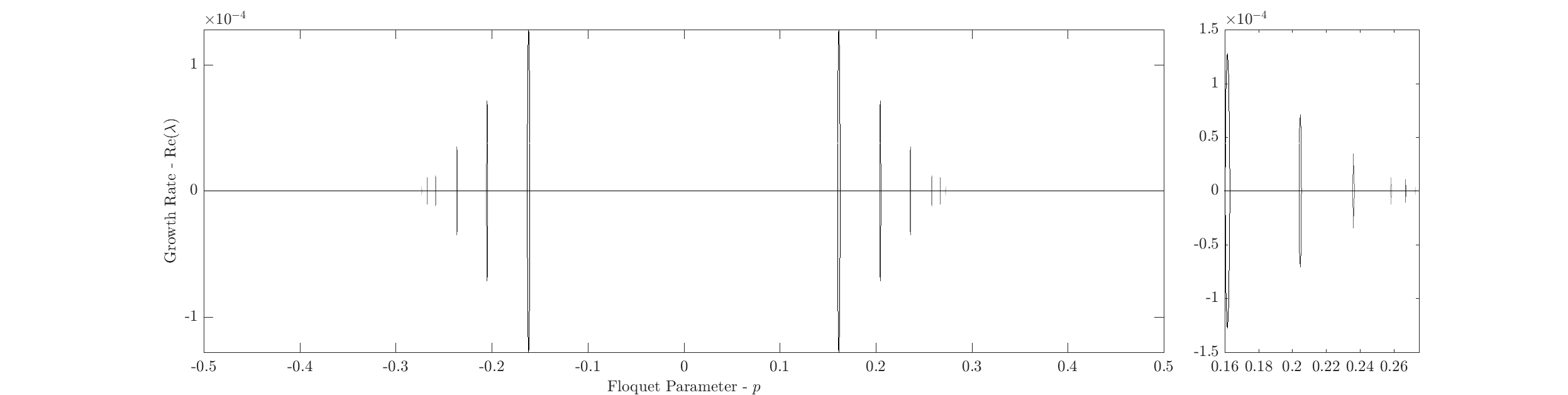}
	\caption{Real part of growth rates as a function of the Floquet parameter $p$ with $\omega_1 = -1$, $\omega_2 =1$, $\rho = 1.028$, $\delta = 1$, $\mu = \sqrt{.1}$ and $\epsilon$ increasing from $\epsilon \approx .001$ (light grey) to $\epsilon \approx .25$ (darker grey). The right panel is a zoomed-in version of a region in the left panel and demonstrates that the spikes seen on the left do indeed represent continuous curves in the Floquet parameter.  As the amplitude increases, the Floquet parameter associated with the dominant instability decreases.}\label{fig:delta_1_w1_m_1_w2_1_instability}
	\end{figure}

\begin{figure}
	\centering
	\includegraphics[width=\textwidth]{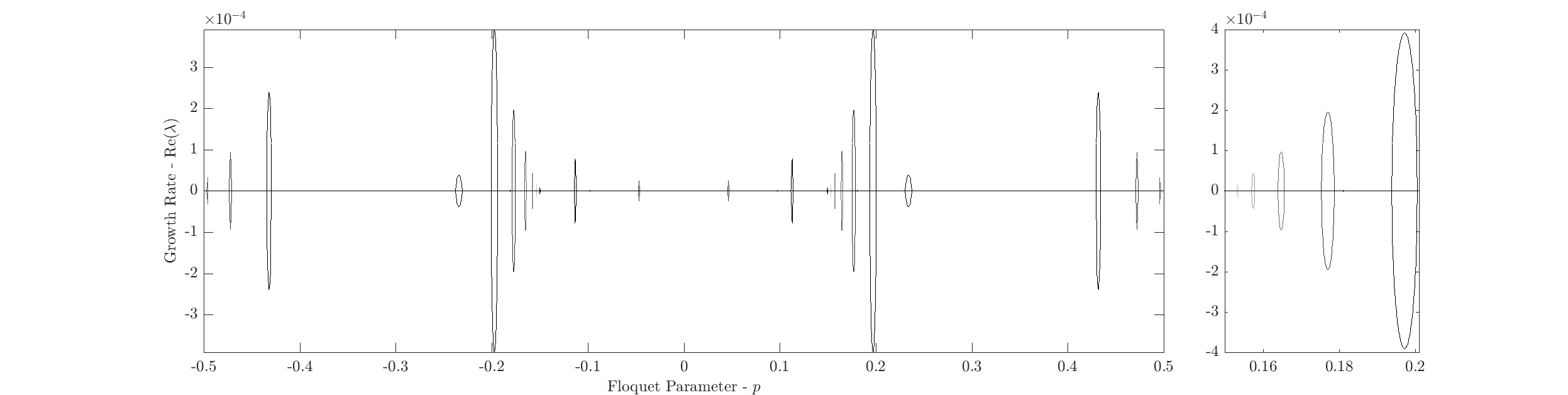}
	\caption{Real part of growth rates as a function of the Floquet parameter $p$ with $\omega_1 = 1$, $\omega_2 =1$, $\rho = 1.028$, $\delta = 4$, $\mu = \sqrt{.1}$ and $\epsilon$ increasing from $\epsilon \approx .001$ (light grey) to $\epsilon \approx .015$ (darker grey).  The right panel is a zoomed-in version of a region in the left panel and demonstrates that the spikes seen on the left do indeed represent continuous curves in the Floquet parameter.  As the amplitude increases, the Floquet parameter associated with the dominant instability increases.}\label{fig:delta_4_w1_1_w2_1_instability}
	\end{figure}

\begin{figure}
	\centering
	\includegraphics[width=\textwidth]{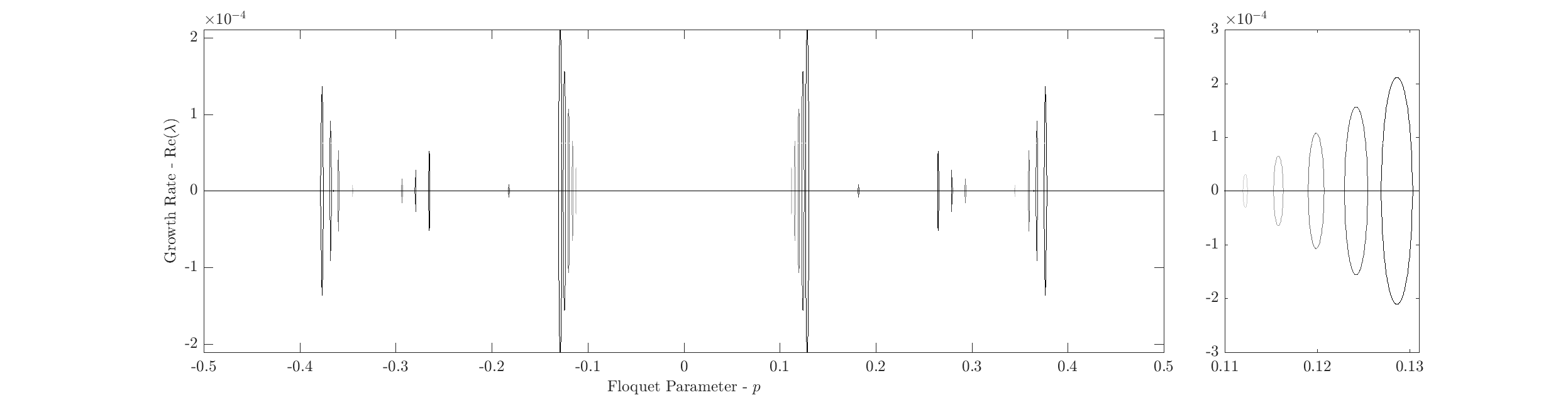}
	\caption{Real part of growth rates as a function of the Floquet parameter $p$ with $\omega_1 = -1$, $\omega_2 =-1$, $\rho = 1.028$, $\delta = 4$, $\mu = \sqrt{.1}$ and $\epsilon$ increasing from $\epsilon \approx .001$ (light grey) to $\epsilon \approx .03$ (darker grey). The right panel is a zoomed-in version of a region in the left panel and demonstrates that the spikes seen on the left do indeed represent continuous curves in the Floquet parameter.  As the amplitude increases, the Floquet parameter associated with the dominant instability increases.}\label{fig:delta_4_w1_m_1_w2_m_1_instability}
	\end{figure}

For simplicity, in Figure \ref{fig:fixw1w2}, we have chosen $\mu = \delta$. While this particular choice is unnecessary, not choosing so further expands the possible parameter space to be explored. In fixing either $\omega_1$ or $\omega_2$ to be zero, we can seen how the threshold between $\delta$ and the strength of the shear must be related in order from small amplitude traveling wave solutions to be susceptible to MI. Figure \ref{fig:fixdelta} shows how this relationship becomes more complicated as both $\omega_1$ and $\omega_2$ are simultaneously varied when $\mu = \delta = 2$.  We note that we do not find MI for example when  $\mu = \sqrt{0.1}$ using the solutions presented in Section 3 (see Figures \ref{fig:delta_1_w1_m_1_w2_1_instability}, \ref{fig:delta_4_w1_1_w2_1_instability}, and \ref{fig:delta_4_w1_m_1_w2_m_1_instability}).  Whether this is a general feature for all solutions, and what the exact parameter values are which ensure the existence of MIs is a subject of future study. 

We note that these results generalize the behavior of both MIs and HFIs in the irrotational case.  These results also expand on those for a single-layer fluid, where the strength of the linear shear has a dramatic effect on the instability of TWSs as discussed in \cite{thomas2012nonlinear} for approximate models, and in \cite{sprenger} for the full water-wave problem.

\begin{rmk} It is worthwhile to note that for small-amplitude solutions, finding these small bands of instability requires taking a dense grid for $p \in \left[0,1/2\right]$ and presents a computational challenge as further refinement yields more bands of instability, though with decreasing magnitudes.  To reduce the computational time and improve resolution, one can exploit the Hamiltonian structure (see \cite{wahlen,trichtchenko}) and use Krein signature arguments to determine values of $p$ for which there is a potential instability.  These ideas, and their extension to multi-layer flow with depth-varying shear will be explored in a future paper.
	\end{rmk}

% = = = = = = = = = = = = = = = = = = = = = = = = = = = = = = = = = = = = = = = = = = = = = = = = 
\section{Concluding Remarks}
% = = = = = = = = = = = = = = = = = = = = = = = = = = = = = = = = = = = = = = = = = = = = = = = = 
Though we focused on internal waves limited to a one-dimensional interface represented by a graph, we  can easily adapt our problem formulation to an arc-length parametric representation such as those outlined in \cite{akers2016overturned}.  Overturning waves, as well as the Hamiltonian structure of our will be explored in a future paper.  Despite our focus on single-valued waves, we show a variety of solution structures depending on the density and depth ratios yielding waves of elevation or depression as characterized by the sign of $c_2$ in our formulation.  This is directly related to the second-order correction of the tangent angle of the interface as prescribed by the vortex-sheet formulation of the interface for irrotational fluids over infinite depth in \cite{akers2016overturned}.  Here, we have extended these results to finite-depth and linear shear.  

For the single-valued solutions, we show that internal waves can be stabilized by the strength of the linear shear in each layer.  While one can amplify the instabilities seen in Figure \ref{fig:BF_w1_m1_w2_1_delta_1}, the modulational instabilities can be fully suppressed to machine precision if the differences in magnitudes of the depths of the stratified layers is large enough (see Figure \ref{fig:fixw1w2} for details).  This ability of shear to both amplify and suppress modulational stability in bistratified fluids generalizes the phenomena seen for the single-layer fluid where modulational instabilities are suppressed provided that the strength of the vorticity exceeds a critical threshold (see \cite{thomas2012nonlinear,sprenger} for details).  

% = = = = = = = = = = = = = = = = = = = = = = = = = = = = = = = = = = = = = = = = = = = = = = = = 
\textbf{Acknowledgements.}  This work was completed at the Institute for Computational and Experimental Research in Mathematics (ICERM) in Providence, RI, during the Spring 2017 semester.  Both KO and CC greatfully acknowledge the support of ICERM and their computing facilities.  KO also acknowledges the support of the National Science Foundation under grant number DMS-1313049. Any opinions, findings, and conclusions or recommendations expressed in this material are those of the authors and do not necessarily reflect the views of the National Science Foundation.

\bibliographystyle{plain}
\bibliography{waves_strat_fluid}
\appendix
% !Tex root = ../rigid_lid_stability_v1.tex
\section{Stability Formulation}
	
	For simplicity, we replace $x-ct$ with $x$ where the non-dimensinoal variables $\eta$ and $q_j$ ($j = 1, 2$) are periodic in the new variable $x$ with period $2\pi$.  
	Substituting the expansions for $\eta$, $q_1$ and $q_2$ into \eqref{eqn:bernoulliNonDim}, we find at $\mathcal{O}(\varepsilon)$

	\begin{multline}
		V \eta_1(x) - T_1\partial_x\alpha_1(x) + T_2\partial_x\beta_1(x) \\= \lambda \left( \bigg(-\Omega \partial_x^{-1} - \left(T_1-T_2\right)(\partial_x\eta_0)\bigg)\eta_1(x) + \alpha_1(x) -\rho\beta_1(x)\right)
		\end{multline}
	where 
		$$ V=\left((\rho-1)+\omega_1T_1-\omega_2T_2+ \epsilon\mu^2\epsilon\mu^2(\partial_x\eta_0)\left(T_1^2 - \frac{1}{\rho}T_2^2\right)\partial_x\right),$$

	$$\Omega = \omega_1 - \rho\omega_2, \qquad T_1 = \frac{\left(\epsilon\alpha_0(x) - \omega_1\epsilon\eta_0(x)-c\right)}{1 + \epsilon^2\mu^2(\partial_x\eta_0)^2},\quad  \text{ and } \quad T_2 = \frac{\rho\:\left(\epsilon\beta_0(x)-\omega_2\epsilon\eta_0(x) - c\right)}{1 +  \epsilon^2\mu^2(\partial_x\eta_0)^2}.$$

	For the nonlocal equations given by \eqref{eqn:nonlocal_top_spatial}, and \eqref{eqn:nonlocal_bottom_spatial}, we find 

	\begin{multline}
		\lambda\left<e^{-ikx}\mathcal{C}_{1,k}\eta_1(x)\right> = \\\left<e^{-ikx}\left(-ik\:\mathcal{C}_{1,k}\left(\epsilon(\partial_x\alpha_0)-\omega_1\epsilon\eta_0-c\right)\:\eta_1(x)\right)\right>-\frac{i}{\mu}\left<e^{-ikx}\left(\mathcal{S}_{1,k}\partial_x \alpha_1(x)\right)\right>,
		\end{multline}
		\begin{multline}
		\lambda\left<e^{-ikx}\mathcal{C}_{2,k}\eta_1(x)\right> = \\\left<e^{-ikx}\left(-ik\:\mathcal{C}_{2,k}\left(\epsilon(\partial_x\beta_0)-\omega_2\epsilon\eta_0(x)-c\right)\:\eta_1(x)\right)\right>-\frac{i}{\mu}\left<e^{-ikx}\left(\mathcal{S}_{2,k}\partial_x \beta_1(x)\right)\right>.\end{multline}

%-------------------------------------------------------------------------------------------------------------
%-------------------------------------------------------------------------------------------------------------
\section{Generalized Matrix Eigenvalue Problem}

Since $\bar\alpha_1$, $\bar\beta_1$ and $\bar\eta_1$ are 2$\pi$ periodic, we can represent the functions by their Fourier Series given by $$\bar\alpha_1(x) =\sum_{n=-\infty}^{\infty}e^{in\pi x} \hat A_n, \quad \bar\beta_1(x) = \sum_{n=-\infty}^{\infty}e^{in\pi x} \hat B_n, \quad \bar\eta_1(x) =\sum_{n=-\infty}^{\infty}e^{in\pi x} \hat{N}_n.$$  Following the steps outlined in Section \ref{sec:stability}, the resulting eigenvalue problem takes the for
\begin{equation}
\begin{bmatrix} 
			\mathcal{L}_{1,1} & \mathcal{L}_{1,2} & \mathcal{L}_{1,3}\\  \mathcal{L}_{2,1} & \mathcal{L}_{2,2} & 0\\ \mathcal{L}_{3,1} & 0 & \mathcal{L}_{3,3} \end{bmatrix} X = \lambda\begin{bmatrix} \mathcal{R}_{1,1} & \mathcal{R}_{1,2} & \mathcal{R}_{1,3}\\  \mathcal{R}_{2,1} & 0 & 0\\ \mathcal{R}_{3,1} & 0 & 0 \end{bmatrix} X
			\end{equation}
			where 
$$\mathcal{L}_{1,1}=\left((\rho-1)+\omega_1T_1-\omega_2T_2- \epsilon\mu(\partial_x\eta_0)\left(\frac{1}{\rho}T_2^2 - T_1^2\right)\partial_x\right),\quad \mathcal{L}_{1,2} = -T_1\partial_x,$$ 
$$ \quad \mathcal{L}_{1,3} = T_2\partial_x, \qquad \mathcal{L}_{2,1}(k) = k\left(\epsilon(\partial_x\alpha_0)- \epsilon\omega_1\eta_0(x)-c\right)\cosh(\mu k(\epsilon\eta_0 - 1)), $$
$$ \mathcal{L}_{2,2}(k) = \frac{1}{\mu}\sinh(\mu k(\epsilon\eta_0-1))\partial_x,\qquad \mathcal{L}_{3,1}(k) = k\left(\epsilon(\partial_x\beta_0) - \epsilon\omega_2\eta_0(x)-c\right)\cosh(\mu k(\epsilon\eta_0 +\delta)),$$ 
$$\mathcal{L}_{3,3}(k) = \frac{1}{\mu}\sinh(\mu k(\epsilon\eta_0+\delta))\partial_x,$$
	and
	$$\mathcal{R}_{1,1} = \Omega\partial_x^{-1}+(T_1-T_2)(\partial_x\eta_0), \qquad \mathcal{R}_{1,2}=1, \qquad \mathcal{R}_{1,3}=-\rho,$$
		$$\mathcal{R}_{2,1}(k) = i\cosh(\mu k(\epsilon\eta_0 - 1)) \qquad \mathcal{R}_{3,1}(k) = i\cosh(\mu k(\epsilon\eta_0 +\delta))$$

	\noindent where $k = p + \ell,$ for $\ell \in \mathbb{Z}$, and  $X = e^{i\mu x}\left[\bar\eta_1~\bar\alpha_1 ~\bar\beta_1 \right]^{{T}}$.

Instead of directly solving the eigenvalue problem as stated above, a more stable approach is to calculate the eigenvalues is to reformulate the problem as a quadratic eigenvalue problem.  Noting that both $\mathcal{L}_{2,2}$ and $\mathcal{L}_{3,3}$ are invertible (\cite{craig2005hamiltonian}), we can formally replace both $\bar{\alpha}_1$ and $\bar\beta_1$ with the expressions 

\begin{equation}
	\bar\alpha_1= \mathcal{L}_{2,2}^{-1}\left(\lambda \mathcal{R}_{2,1} - \mathcal{L}_{2,1}\right)\bar\eta_1, \quad 
	\bar\beta_1= \mathcal{L}_{3,3}^{-1}\left(\lambda \mathcal{R}_{3,1} - \mathcal{L}_{3,1}\right)\bar\eta_1
	\end{equation}

Using these expressions in the $3\times 3$ eigenvalue problem above, we can reduce the system to a quadratic eigenvalue problem of the form 
\begin{equation}
	\left(\lambda^2\mathcal{A}_2  + \lambda \mathcal{A}_1 + \mathcal{A}_0\right)\bar\eta_1 = 0,
	\end{equation}
	where the expressions for the operators $\mathcal{A}_j$'s are given by 
\begin{eqnarray}
	\mathcal{A}_0 &=& \mathcal{L}_{1,1} - \mathcal{L}_{1,2}\mathcal{L}_{2,2}^{-1}\mathcal{L}_{2,1} - \mathcal{L}_{1,3}\mathcal{L}_{3,3}^{-1}\mathcal{L}_{3,1},\\
	\mathcal{A}_1 &=& -\mathcal{R}_{1,1} + \mathcal{R}_{1,2}\mathcal{L}_{2,2}^{-1}\mathcal{L}_{2,1} + \mathcal{L}_{1,2}\mathcal{L}_{2,2}\mathcal{R}_{2,1} + \mathcal{L}_{1,2}\mathcal{L}_{3,3}^{-1}\mathcal{R}_{3,1} + \mathcal{R}_{1,3}\mathcal{L}_{3,3}^{-1}\mathcal{L}_{3,1},\\
	\mathcal{A}_2 &=& -\mathcal{R}_{1,2}\mathcal{L}_{2,2}^{-1}\mathcal{R}_{2,1} - \mathcal{R}_{1,3}\mathcal{L}_{3,3}^{-1}\mathcal{R}_{3,1}.
	\end{eqnarray}

\end{document}